\documentclass[a4paper,fleqn]{cas-dc}

\makeatletter\def\Hy@Warning#1{}\makeatother

\renewcommand{\printorcid}{}
\ExplSyntaxOn
\keys_set:nn { stm / mktitle } { nologo }
\ExplSyntaxOff

\usepackage[numbers]{natbib}

\usepackage{graphicx}
\usepackage{float}
\usepackage{hyperref}
\hypersetup{colorlinks,allcolors=black}
\usepackage{multirow}
\usepackage{todonotes}
\usepackage[normalem]{ulem}
\useunder{\uline}{\ul}{}
\usepackage{subcaption}
\usepackage{url}

\usepackage{lipsum}

\begin{document}
\let\WriteBookmarks\relax
\def\floatpagepagefraction{1}
\def\textpagefraction{.001}

\shorttitle{SHAMSUL: \textbf{S}ystematic \textbf{H}olistic \textbf{A}nalysis to investigate \textbf{M}edical \textbf{S}ignificance \textbf{U}tilizing \textbf{L}ocal interpretability methods}

\shortauthors{Mahbub Ul Alam et al. (mahbub@dsv.su.se; mahbub.ul.alam.anondo@gmail.com)}

\title[mode = title]{SHAMSUL: Systematic Holistic Analysis to investigate Medical Significance Utilizing Local interpretability methods in deep learning for chest radiography pathology prediction}

\tnotemark[1]

\tnotetext[1]{The acronym SHAMSUL, derived from a Semitic word meaning "the Sun," serves as a symbolic representation of our heatmap score-based interpretability analysis approach aimed at unveiling the medical significance inherent in the predictions of black box deep learning models.}

\author{Mahbub Ul Alam} 
\fnmark[a]
\cormark[1]
\cortext[1]{Corresponding author} 
\ead{mahbub@dsv.su.se} 

\author{Jaakko Hollmén}
\fnmark[a]
\ead{jaakko.hollmen@dsv.su.se} 

\author{Jón Rúnar Baldvinsson}
\fnmark[b]
\ead{jon.r.baldvinsson@gmail.com} 

\author{Rahim Rahmani}
\fnmark[a]
\ead{rahim@dsv.su.se}

\address[a]{Department of Computer and Systems Sciences, Stockholm University, Stockholm, Sweden}
\address[b]{Skatturinn (Iceland Revenue and Customs), Reykjavík, Iceland}

\begin{abstract}[S U M M A R Y]

\textbf{The published version of this article can be obtained using the following link:}\\\\\href{https://doi.org/10.5617/nmi.10471}{\underline{https://doi.org/10.5617/nmi.10471}}\\\\
The interpretability of deep neural networks has become a subject of great interest within the medical and healthcare domain. This attention stems from concerns regarding transparency, legal and ethical considerations, and the medical significance of predictions generated by these deep neural networks in clinical decision support systems. To address this matter, our study delves into the application of four well-established interpretability methods: Local Interpretable Model-agnostic Explanations (LIME), Shapley Additive exPlanations (SHAP), Gradient-weighted Class Activation Mapping (Grad-CAM), and Layer-wise Relevance Propagation (LRP). Leveraging the approach of transfer learning with a multi-label-multi-class chest radiography dataset, we aim to interpret predictions pertaining to specific pathology classes. Our analysis encompasses both single-label and multi-label predictions, providing a comprehensive and unbiased assessment through quantitative and qualitative investigations, which are compared against human expert annotation. Notably, Grad-CAM demonstrates the most favorable performance in quantitative evaluation, while the LIME heatmap score segmentation visualization exhibits the highest level of medical significance. Our research underscores both the outcomes and the challenges faced in the holistic approach adopted for assessing these interpretability methods and suggests that a multimodal-based approach, incorporating diverse sources of information beyond chest radiography images, could offer additional insights for enhancing interpretability in the medical domain.
  
\end{abstract}
 

\maketitle
 
\section{Introduction}

Machine learning (ML) \cite{bishop2006pattern} has emerged as a field utilizing probabilistic computational methods to enable machines to learn from examples. ML models, for instance, can be trained to diagnose patients by leveraging labeled medical data \cite{bakator2018deep}. To achieve accurate predictions, these models require fine-tuning of certain parameter values, a process referred to as 'learning the model' \cite{molnar2022}.

Deep Learning (DL) \cite{goodfellow2016deep}, a subfield of ML, enables models to learn intricate representations of data by progressively synthesizing lower-level information into higher-level abstractions. This process minimizes the need for human-engineered features. Deep neural networks (DNNs) are commonly employed in this regard, as they can learn the necessary features from input data through complex compositions of numerous tuning parameters.

The remarkable success of deep learning-based models presents promising opportunities in the healthcare and medicine sectors. Specifically, the medical diagnosis field can benefit from employing deep learning methods as aids for human specialists \cite{ching2018opportunities}. However, the use of deep learning methods in critical environments necessitates transparency in the decision-making process. One major challenge with DNNs is their lack of interpretability in this complex decision-making process. To address this, various interpretability methods have been proposed \cite{e23010018, samek2019explainable}. In the domain of ML/DL, interpretability methods aim to explain the behavior of models, enabling humans to understand specific decisions to a certain extent \cite{miller2019explanation, kim2016examples}. Interpretable models facilitate comprehension of outcomes, thereby aiding evaluations of ethics, trust, causality, and fairness \cite{lipton2018mythos}.

Interpretability plays a crucial role in assessing these criteria, making it imperative to include interpretability in the work involving deep learning models from legal and ethical perspectives. Regulatory frameworks such as the European Union (EU) General Data Protection Regulation (GDPR) \cite{voigt2017eu}, EU Artificial Intelligence Act \cite{veale2021demystifying}, and the US Algorithmic Accountability Act of 2022 \cite{mokander2022us} all emphasize the importance of interpretability.

Interpretability methods can be categorized in various ways. They can be classified as local or global based on the scope of their interpretation \cite{buhrmester2021analysis}. Local methods interpret individual predictions, while global methods interpret all predictions of a model simultaneously. Another classification \cite{ras2022explainable} categorizes interpretability methods as follows: (a) Visualization methods, which highlight specific input features that significantly influence a particular prediction; (b) Model distillation, employing a relatively simple surrogate model to explain the predictions of the main model, such as a DNN; and (c) Intrinsic methods, utilizing specialized DNNs that provide interpretations for predictions. If an interpretability method focuses on a particular kind or category of ML models, it is considered model-specific; other-wise, it is regarded as model-agnostic \cite{molnar2022}.

When using interpretability methods on models trained for medical diagnosis, it is crucial to focus on interpreting the medical significance. Proper consideration of this significance ensures the usability of such models in healthcare sectors, including clinical decision support systems \cite{souza2021quality, xiao2018opportunities}. The medical significance of symptoms or disease processes pertains to their clinical manifestations, which can disrupt the normal functioning of cells, tissues, organs, or organ systems \cite{goodman2014felson}.

Radiography images serve as primary diagnostic tools and enable the analysis of medical significance by evaluating the distinctive shapes and colors in the images \cite{goodman2014felson}. Given the widespread usage of radiography, abundant data is available, making it a suitable choice for DL models \cite{ccalli2021deep} and their interpretability aspects \cite{salahuddin2022transparency}.

In this study, we introduce SHAMSUL, an approach that stands for \textbf{S}ystematic \textbf{H}olistic \textbf{A}nalysis to investigate \textbf{M}edical \textbf{S}ignificance \textbf{U}tilizing \textbf{L}ocal interpretability methods. Our primary focus is to analyze the medical significance of specific pathology classes using a DNN model trained on CheXpert \cite{irvin2019chexpert}, a multi-label multi-class dataset comprising chest radiography images. This DNN model employs transfer learning on a deep convolutional neural network (CNN) architecture \cite{lecun2015deep}.

Our work extends our previous research \cite{9866967} and makes the following contributions:

\begin{enumerate}

\item In addition to the previously used local interpretability methods, Gradient-weighted Class Activation Mapping (Grad-CAM) \cite{selvaraju2017grad} and Layerwise-Relevance Propagation (LRP) \cite{bach2015pixel}, we incorporate two widely-used interpretability methods: Local Interpretable Model-agnostic Explanations (LIME) \cite{ribeiro2016should} and Shapley Additive exPlanations (SHAP) \cite{lundberg2017unified}. We employ these methods to investigate the medical significance of selected pathology classes using heatmap score visualization.

\item We present a meticulous and comprehensive quantitative evaluation of these interpretability methods through an assessment of their performance across diverse pathology classes. In particular, we undertake a comparative analysis by juxtaposing the heatmap score performance of these methods with human expert annotation.

\item Moreover, in order to provide a comprehensive understanding of interpretability, we extensively delve into the qualitative aspects of performance within two selected pathology classes. By conducting a thorough overview, we offer an in-depth analysis of the qualitative attributes of the interpretability methods. In particular, we focus on the instance that achieves the highest quantitative interpretability performance within each selected class, thereby highlighting its significance. Additionally, to ensure methodological consistency, systematic analysis, and impartiality, we also select the best predicted instance based on probability score for each class, as well as one multi-labelled instance.

\end{enumerate}

In conducting this holistic analysis, our goal is to provide a detailed and even-handed perspective on the functionality of these interpretability methods. We endeavor to illuminate the foundational logic driving both effective and ineffective predictions, delving into the theoretical underpinnings, limitations, and prospective insights of these quantitative and qualitative evaluation approaches.

The code implementation of SHAMSUL can be accessed via the following link (accessed on 22 August 2023):\\\href{https://github.com/anondo1969/SHAMSUL}{https://github.com/anondo1969/SHAMSUL}

\section{Related Works} \label{related_works}

This section provides a concise overview of the interpretability methods mentioned earlier and their application in the context of chest radiography images and related biomedical image-based prediction tasks.

LIME has been employed for COVID-19 prediction tasks using chest radiography images \cite{ahsan2020covid}. Additionally, LIME and Grad-CAM have been utilized to investigate the impact of lung segmentation on the prediction of COVID-19 in chest radiography images \cite{teixeira2021impact}. DeepCOVIDExplainer utilized gradient-guided class activation maps (Grad-CAM++) and layer-wise relevance propagation (LRP) \cite{karim2020deepcovidexplainer}. SHAP has been employed to explain the prediction of choroidal neovascularization (CNV), diabetic macular edema (DME), and drusens using optical coherence tomography (OCT) scans \cite{singh2020interpretation}. It has also been used for COVID-19 prediction in chest radiography images \cite{9576766}. CAM (class activation mappings) heatmaps have been utilized to interpret predictions for pneumonia detection \cite{rajpurkar2017chexnet} and multi-pathology detection \cite{rajpurkar2018deep}. The abnormality in chest radiography images was investigated using Grad-CAM \cite{hsu2019development}. The visualization of tuberculosis was explored using saliency maps and Grad-CAM \cite{pasa2019efficient}. LRP has been employed in MRI-based Alzheimer's disease classification \cite{bohle2019layer} task. In \cite{van2020systematic}, the authors conducted an investigation into the comparative performances of selected saliency methods, including LRP and Grad-CAM, utilizing heatmaps in the context of diabetic retinopathy lesion detection as a benchmark task. The quantitative evaluation of Grad-CAM, LRP, and five additional saliency methods was conducted using chest X-ray data \cite{saporta2022benchmarking}. It is worth noting that these works did not perform a comprehensive comparison of the four interpretability methods, as previously mentioned.

\section{Methods and the Dataset}

In this section, we present the details of the interpretability methods and datasets utilized in SHAMSUL.

\subsection{LIME}

One of the interpretability methods employed in SHAMSUL is LIME, which stands for Local Interpretable Model-agnostic Explanations \cite{ribeiro2016should}. LIME operates under the assumption that the local behavior of a model is simpler to compute than its global behavior. Based on this assumption, LIME approximates the local behavior of the model to provide explanations for individual predictions. This method can be applied to various types of inputs, including images, tabular data, and text.

LIME focuses on explaining the prediction for a single instance by considering its local neighborhood. It expands the input data around the specific instance to gather more information about the local area. Subsequently, LIME employs a sparse linear model, such as the least absolute shrinkage and selection operator (lasso) \cite{tibshirani1996regression}, to train on the augmented data. By using a linear model, LIME ensures interpretability, as linear models are inherently more interpretable. The regularization factor in the linear model is fine-tuned to obtain a suitable number of non-zero coefficients, facilitating the interpretation of the specific input.

To assign weights to the surrounding points in the local neighborhood, LIME employs a kernel function \cite{bergman1951kernel}, with the default choice being a radial basis function kernel. The weights are calculated based on the distance between the points and the instance under consideration.

\begin{equation}
\xi (y) = argmin_{g \in G} L(f,g,\pi_{x'}) + \Omega (g)
\label{lime_method}
\end{equation} 

The mathematical representation of the LIME explanation for a particular input $x$ is provided by Equation \ref{lime_method}. Let $f$ denote the trained model, and $g$ represent the linear surrogate model. The proximity area of $x$ is denoted as $\pi_{x'}$, where $x'$ represents the surrounding points. $\Omega(g)$ represents the penalty estimation for the complexity of $g$. The aim of LIME is to minimize the locality-aware squared case, denoted by $L$.

\subsection{SHAP}

The SHAP (Shapley Additive exPlanations) method \cite{lundberg2017unified} is employed in this study to provide interpretability for individual predictions. Consider a model $f : X \rightarrow \mathbb{R}$, where we aim to interpret the prediction $f(x)$ for a given instance $x \in X$. Let $M$ denote the set of features present in $x$. The binary vector $z \in {0, 1}^M$ represents the absence or presence of each feature. By removing the features set to $0$ in $z$ from $x$, we obtain an input sample denoted as $h(z) \in X$.

\begin{equation}
\varphi_i(x) = \sum_{ z{ \in \{0, 1\}^M} \setminus\{i\} } \frac{ \mid z \mid!( M-\mid z \mid -1 )! }{ M! } B
\label{shap_method}
\end{equation}

\begin{equation}
B=(f(h(z \cup \{i\} )) -f(h(z)))
\end{equation}

The attribution score, determined by the Shapley value, can be calculated for each feature $1 \leq i \leq M$. Here, the cardinality operator $\mid \cdot \mid$ represents the number of non-zero elements in a set. To approximate the Shapley values in our work, we utilized KernelSHAP \cite{lundberg2017unified}, which combines the concepts of LIME and Shapley values. In KernelSHAP, a linear model $g : \mathbb{R}^M \rightarrow \mathbb{R}$ is trained using a weighted least squares (WLS) regression method, with its optimal coefficients serving as the Shapley values. The data points are uniformly sampled from the simplified input space ${0, 1}^M$, considering the outputs as the value of $f \circ h$.

\subsection{Grad-CAM}

Grad-CAM (Gradient-weighted Class Activation Mapping) \cite{selvaraju2017grad} is a technique that assesses the significance of individual neurons in an input using a feature map. This feature map is generated by leveraging the gradient scores propagated to the final convolutional layer of a Convolutional Neural Network (CNN). By utilizing Grad-CAM, it becomes possible to produce a heatmap that highlights the salient features of the input.

Let $y_c$ denote the class score for a specific class $c$. Each neuron's significance can be interpreted as $y_c$ and is derived by calculating the gradient of $y_c$ with respect to the activation $A$. This process yields an importance weight $\alpha_k^c$ for each feature map $k$, given by:

\begin{equation}
	\alpha_k ^c = \frac{1}{z} \sum_i \sum_j \frac{\partial y_c}{\partial A_{ij} ^k}
\end{equation}

The heatmap can now be generated as a weighted combination of the activation maps. To emphasize positive cases exclusively, a Rectified Linear Unit (ReLU) function \cite{agarap2018deep} is employed to discard negative values. Hence, the Grad-CAM heatmap $L_{Grad-CAM}^c$ is defined as:

\begin{equation}
	L_{Grad-CAM}^c = ReLU(\sum_k \alpha_k ^c A^k)
\end{equation}

This computational process yields the Grad-CAM heatmap, effectively highlighting the important features within the input that are associated with the class $c$.

\subsection{LRP}

LRP (Layerwise-Relevance Propagation) \cite{bach2015pixel} operates under the assumption that a model can be decomposed into multiple layers of computation. It leverages the backpropagation algorithm \cite{rumelhart1986learning} to calculate the importance of each neuron, propagating from the final layer to the input layer while traversing through the intermediate layers. The relevance of each neuron $R_i^l$ can be computed using the following equation:

\begin{equation}
R_i^l= \sum_j R_{i\leftarrow j} ^{(l,l+1)} = \sum_j \frac{z_{ij}}{z_j} R_j ^{l+1}
\label{lrp_method}
\end{equation}

This equation represents the relevance $R_i^l$ of a neuron in a lower layer, based on the relevance $R_j^{l+1}$ of the corresponding neuron in the upper layer. The localized pre-activations are denoted as $z_{ij} = x_i w_{ij}$, where $x_i$ represents the neuron's input and $w_{ij}$ signifies the weight associated with the connection. Additionally, $z_j= \sum_i z_{ij}+b_j$, where $b_j$ denotes the bias term. Through this layerwise-relevance propagation mechanism, LRP enables the assessment of the importance of neurons throughout the network architecture.

\begin{table}[t]
\caption{Classification details of the CheXpert dataset \cite{irvin2019chexpert}.}
\label{chexpert_data_details}
\begin{tabular}{|l|l|l|l|}
\hline
\textbf{Pathology} & \begin{tabular}[c]{@{}l@{}}\textbf{Positive}\\(\textbf{\%})\end{tabular} & \begin{tabular}[c]{@{}l@{}}\textbf{Uncertain}\\(\textbf{\%})\end{tabular} & \begin{tabular}[c]{@{}l@{}}\textbf{Negative}\\(\textbf{\%})\end{tabular} \\ \hline
\begin{tabular}[c]{@{}l@{}}No\\Finding\end{tabular} & \begin{tabular}[c]{@{}l@{}}16627\\(8.86)\end{tabular} & \begin{tabular}[c]{@{}l@{}}0\\ (0.0)\end{tabular} & \begin{tabular}[c]{@{}l@{}}171014\\(91.14)\end{tabular} \\ \hline
\begin{tabular}[c]{@{}l@{}}Enlarged\\Cardiomediastinum\end{tabular} & \begin{tabular}[c]{@{}l@{}}9020\\ (4.81)\end{tabular} & \begin{tabular}[c]{@{}l@{}}10148\\(5.41)\end{tabular} & \begin{tabular}[c]{@{}l@{}}168473\\(89.78)\end{tabular} \\ \hline
Cardiomegaly & \begin{tabular}[c]{@{}l@{}}23002\\(12.26)\end{tabular} & \begin{tabular}[c]{@{}l@{}}6597\\(3.52)\end{tabular} & \begin{tabular}[c]{@{}l@{}}158042\\(84.23)\end{tabular} \\ \hline
\begin{tabular}[c]{@{}l@{}}Lung\\Lesion\end{tabular} & \begin{tabular}[c]{@{}l@{}}6856\\(3.65)\end{tabular} & \begin{tabular}[c]{@{}l@{}}1071\\(0.57)\end{tabular} & \begin{tabular}[c]{@{}l@{}}179714\\(95.78)\end{tabular} \\ \hline
\begin{tabular}[c]{@{}l@{}}Lung\\Opacity\end{tabular} & \begin{tabular}[c]{@{}l@{}}92669\\(49.39)\end{tabular} & \begin{tabular}[c]{@{}l@{}}4341\\(2.31)\end{tabular} & \begin{tabular}[c]{@{}l@{}}90631\\(48.3)\end{tabular} \\ \hline
Edema & \begin{tabular}[c]{@{}l@{}}48905\\(26.06)\end{tabular} & \begin{tabular}[c]{@{}l@{}}11571\\(6.17)\end{tabular} & \begin{tabular}[c]{@{}l@{}}127165\\(67.77)\end{tabular} \\ \hline
Consolidation & \begin{tabular}[c]{@{}l@{}}12730\\(6.78)\end{tabular} & \begin{tabular}[c]{@{}l@{}}23976\\(12.78)\end{tabular} & \begin{tabular}[c]{@{}l@{}}150935\\(80.44)\end{tabular} \\ \hline
Pneumonia & \begin{tabular}[c]{@{}l@{}}4576\\(2.44)\end{tabular} & \begin{tabular}[c]{@{}l@{}}15658\\(8.34)\end{tabular} & \begin{tabular}[c]{@{}l@{}}167407\\(89.26)\end{tabular} \\ \hline
Atelectasis & \begin{tabular}[c]{@{}l@{}}29333\\(15.63)\end{tabular} & \begin{tabular}[c]{@{}l@{}}29377\\(15.66)\end{tabular} & \begin{tabular}[c]{@{}l@{}}128931\\(68.71)\end{tabular} \\ \hline
Pneumothorax & \begin{tabular}[c]{@{}l@{}}17313\\(9.23)\end{tabular} & \begin{tabular}[c]{@{}l@{}}2663\\(1.42)\end{tabular} & \begin{tabular}[c]{@{}l@{}}167665\\(89.26)\end{tabular} \\ \hline
\begin{tabular}[c]{@{}l@{}}Pleural\\Effusion\end{tabular} & \begin{tabular}[c]{@{}l@{}}75696\\(40.34)\end{tabular} & \begin{tabular}[c]{@{}l@{}}9419\\(5.02)\end{tabular} & \begin{tabular}[c]{@{}l@{}}102526\\(54.64)\end{tabular} \\ \hline
\begin{tabular}[c]{@{}l@{}}Pleural\\Other\end{tabular} & \begin{tabular}[c]{@{}l@{}}2441\\(1.3)\end{tabular} & \begin{tabular}[c]{@{}l@{}}1771\\(0.94)\end{tabular} & \begin{tabular}[c]{@{}l@{}}183429\\(97.76)\end{tabular} \\ \hline
Fracture & \begin{tabular}[c]{@{}l@{}}7270\\(3.87)\end{tabular} & \begin{tabular}[c]{@{}l@{}}484\\(0.26)\end{tabular} & \begin{tabular}[c]{@{}l@{}}179887\\(95.87)\end{tabular} \\ \hline
\begin{tabular}[c]{@{}l@{}}Support\\Devices\end{tabular} & \begin{tabular}[c]{@{}l@{}}105831\\(56.4)\end{tabular} & \begin{tabular}[c]{@{}l@{}}898\\(0.48)\end{tabular} & \begin{tabular}[c]{@{}l@{}}80912\\(43.12)\end{tabular} \\ \hline
\end{tabular}
\end{table}

\subsection{CheXpert Dataset}

The CheXpert (\textbf{Che}st e\textbf{Xpert}) \cite{irvin2019chexpert} dataset, developed by researchers at Stanford University, comprises a collection of chest X-ray studies from patients who visited Stanford Hospital's inpatient and outpatient facilities from October 2002 to July 2017. Its use is restricted to non-clinical, non-commercial research, as specified by its licensing terms. Publicly available in large (440 GB) and small (11 GB) versions, the dataset images have reduced grayscale levels and, in the small version, decreased resolution to roughly 390 × 320 pixels. In this study, the small version is utilized for the pathology prediction task.

This dataset is carefully devoid of any private or personally identifiable information, with images containing protected health information (PHI) being excluded. The Stanford Hospital Institutional Review Board (IRB) approved the dataset, which was created without individual patient notifications. Stanford Hospital's policy informs patients of potential data sharing for research, with individual consent waived for de-identified data as per IRB and federal guidelines. The dataset lacks a mechanism for patient data removal, although Stanford Hospital allows patients to opt out of future research \cite{garbin2021structured}. 

The dataset includes a training set with 223,414 images from 64,540 patients, and a validation set with 234 images from 200 patients. Each patient's data is exclusively in one set. Each image is annotated with labels for fourteen pathology observations. Table \ref{chexpert_data_details} provides a summary of the dataset's characteristics. It is annotated with three distinct labels: \textit{positive}, \textit{negative}, and \textit{uncertain}, reflecting the presence, absence, or ambiguity of pathologies. In the training set, labels were extracted from X-ray reports’ Impression sections using a rule-based labeler and a phrase list curated by multiple board-certified radiologists. To enhance accuracy, the labeler was developed with a carefully curated phrase list and evaluated against 1000 manually processed radiology reports by two board-certified radiologists. For the validation set, three board-certified radiologists independently annotated each study. Observations were categorized as present, uncertain likely, uncertain unlikely, or absent, and then binarized for analysis. The majority annotation serves as the ground truth, based on image assessments, not reports \cite{garbin2021structured}.

\subsection{CheXlocalize Dataset}

The CheXlocalize \cite{saporta2022benchmarking} dataset serves as a valuable resource for the localization of 10 distinct pathologies on chest X-rays, offering radiologist-annotated segmentations. This dataset encompasses two types of radiologist annotations: (1) pixel-level segmentations and (2) most-representative points. These annotations were meticulously applied to images extracted from the two different CheXpert data sets. The first set comprises 234 chest X-rays derived from 200 patients, while the second set comprises 668 chest X-rays obtained from 500 patients. The chest X-rays were manually segmented by a pair of radiologists, both of whom hold board certifications and possess substantial professional experience, spanning 18 and 27 years respectively. This segmentation was facilitated through the use of the annotation software, MD.ai (\href{https://www.md.ai/}{\underline{https://www.md.ai/}}). The task assigned to these medical experts involved delineating the regions of interest corresponding to each observation within the chest X-rays that had a positive ground-truth label, as identified in the CheXpert dataset \cite{irvin2019chexpert}.

\section{Experimental Setup} \label{experimental_setup}

\subsection{Pathology Prediction}

The experiment setup for our current study remained consistent with our previous work on prediction and visualization \cite{9866967}, utilizing the same trained model. The CheXpert dataset was employed, employing the provided train-test split (223414 train data instances and 234 test data instances). The task at hand involved multi-label classification for pathology prediction, wherein the \textit{uncertain} labels were treated as negative instances. To achieve this, a DenseNet-121 (Dense Convolutional Network) \cite{huang2017densely} was adopted as the base architecture for transfer learning, leveraging pre-trained weights from ImageNet \cite{deng2009imagenet}. For compatibility with the DenseNet-121 model, the input chest radiography images were resized to $224 \times 224$ pixels. Subsequently, the images were scaled by a factor of 255 and converted to RGB (red-green-blue) values. The backpropagation algorithm employed was the 'Adam optimizer', with a minibatch size of 16 and training conducted for 100 epochs.

We used AUROC and AUPRC metrics for the pathology prediction evaluation score. The AUROC (area under the receiver operating characteristic) metric considers the trade-off between the true-positive rate (TPR) and the false-positive rate (FPR) across various decision thresholds. Additionally, we present the weighted average AUROC and AUPRC (area under the precision-recall curve) scores, encompassing all classes except for the \textit{Fracture} class, which lacks positive instances in the test data. The precision-recall curve, which illustrates the relationship between precision (y-axis) and recall (x-axis) at various decision thresholds, is a valuable metric for evaluating the performance of a classification model. It provides insights into the trade-off between correctly identified positive instances (precision) and the model's ability to capture all positive instances (recall). \cite{davis2006relationship}.

\subsection{Quantitative Interpretability Evaluation}

In the application of various interpretability methods, the standard configurations were employed without modification. Specifically, for both LIME and SHAP, the sample size was consistently set at 1000 for each method.

In order to quantitatively evaluate and visualize the obtained relevance from the interpretability methods, pixel-level segmentations were employed. These segmentations were generated by converting the interpretability heatmap score values, where high or positive scores were indicative of the relevance or significance of specific pathology classes within the corresponding pixel regions. The methodology employed for generating segmentations was based on the approach outlined in \cite{saporta2022benchmarking}. To create binary segmentations, a threshold was applied to each heat map. The determination of thresholding for each pathology was conducted using Otsu's method \cite{otsu1979threshold}. This method involves an iterative search for a threshold value that optimally maximizes the variance in pixel intensity between different classes.

To assess the quality of the interpretability methods' generated segmentations, we utilized the mean Intersection over Union (IoU) metric \cite{saporta2022benchmarking}. The IoU measures the overlap between the predicted segmentations from the interpretability methods and the ground-truth annotation segmentations. This metric quantifies the ratio between the area of overlap and the area of union, resulting in a value between 0 and 1. A value of 0 signifies no overlap, while a value of 1 indicates perfectly aligned segmentations. By employing this metric, we can evaluate the extent to which the interpretability methods' generated segmentations align with the human expert annotation segmentations, providing insight into the accuracy and effectiveness of the interpretability methods.

\subsection{Qualitative Interpretability Analysis}

\begin{figure}[ht]
    \centering
    \begin{subfigure}{.18\textwidth}
        \centering
        \includegraphics[width=1\linewidth]{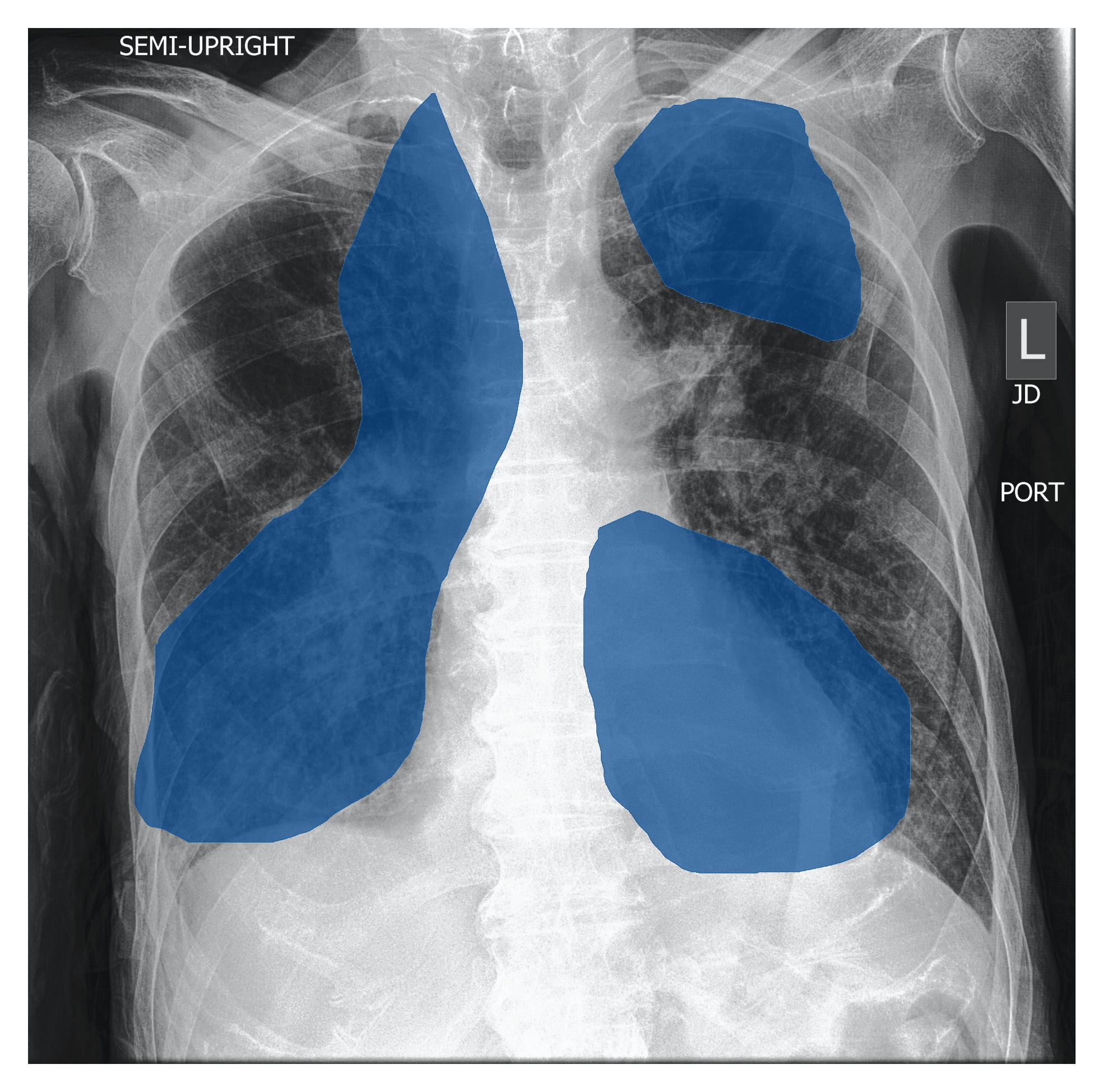}
        \caption{\textit{Lung Opacity} }
        \label{gt_single_prob_Lung_Opacity}
    \end{subfigure} 
    \begin{subfigure}{.18\textwidth}
        \centering
        \includegraphics[width=1\linewidth]{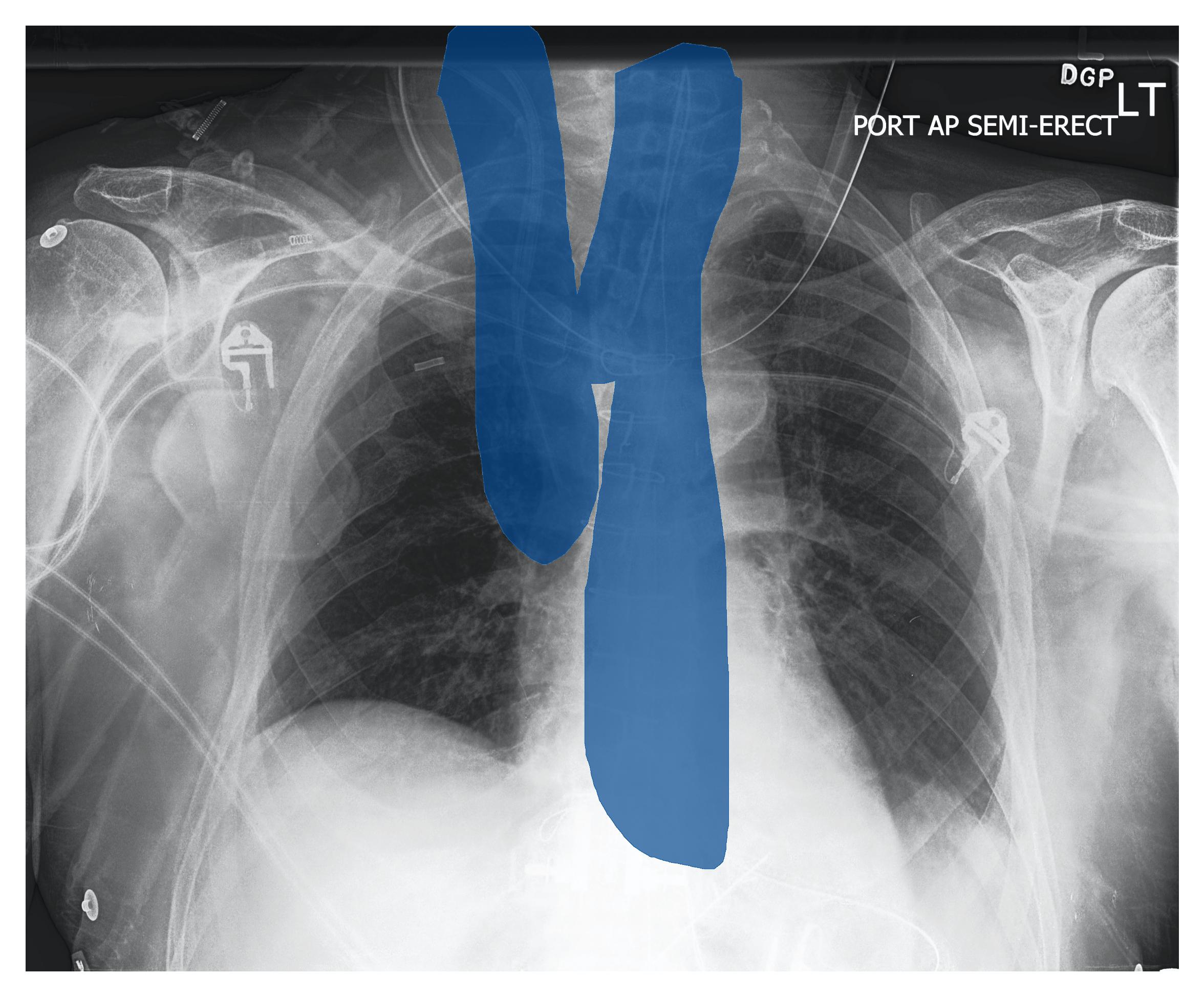}
        \caption{\textit{Support Devices} }
        \label{gt_single_prob_Support_Devices}
    \end{subfigure}

    \caption{Instances of the \textit{Lung Opacity} (prediction probability score: 0.63) and \textit{Support Devices} (prediction probability score: 0.93) classes with human expert segmentation annotation for subsequent qualitative analysis.}
    
    \label{gt_single_prob}
\end{figure}

\begin{figure}[ht]
    \centering
    \begin{subfigure}{.18\textwidth}
        \centering
        \includegraphics[width=1\linewidth]{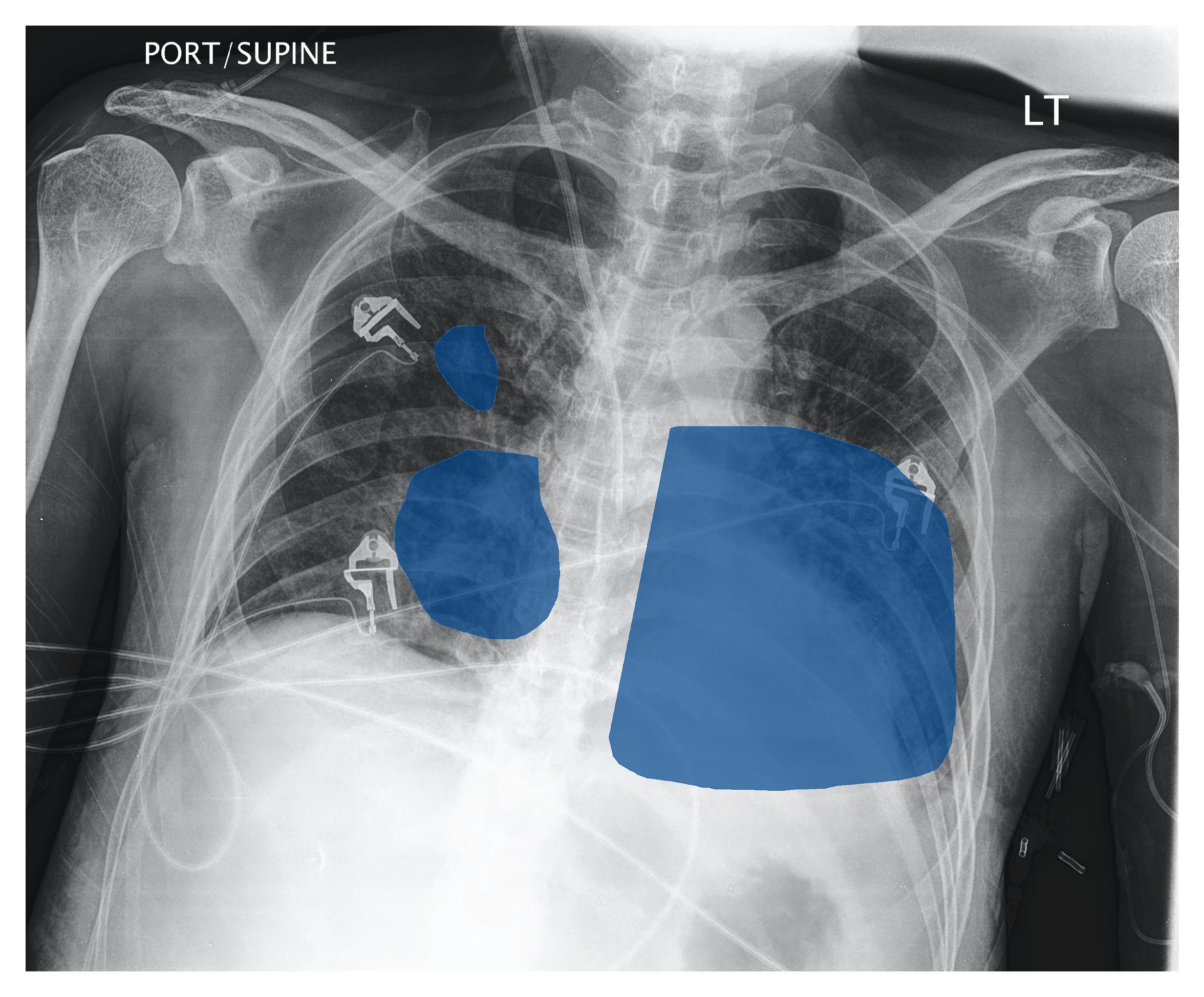}
        \caption{\textit{Lung Opacity} }
        \label{gt_multi_label_prob_Lung_Opacity}
    \end{subfigure} 
    \begin{subfigure}{.18\textwidth}
        \centering
        \includegraphics[width=1\linewidth]{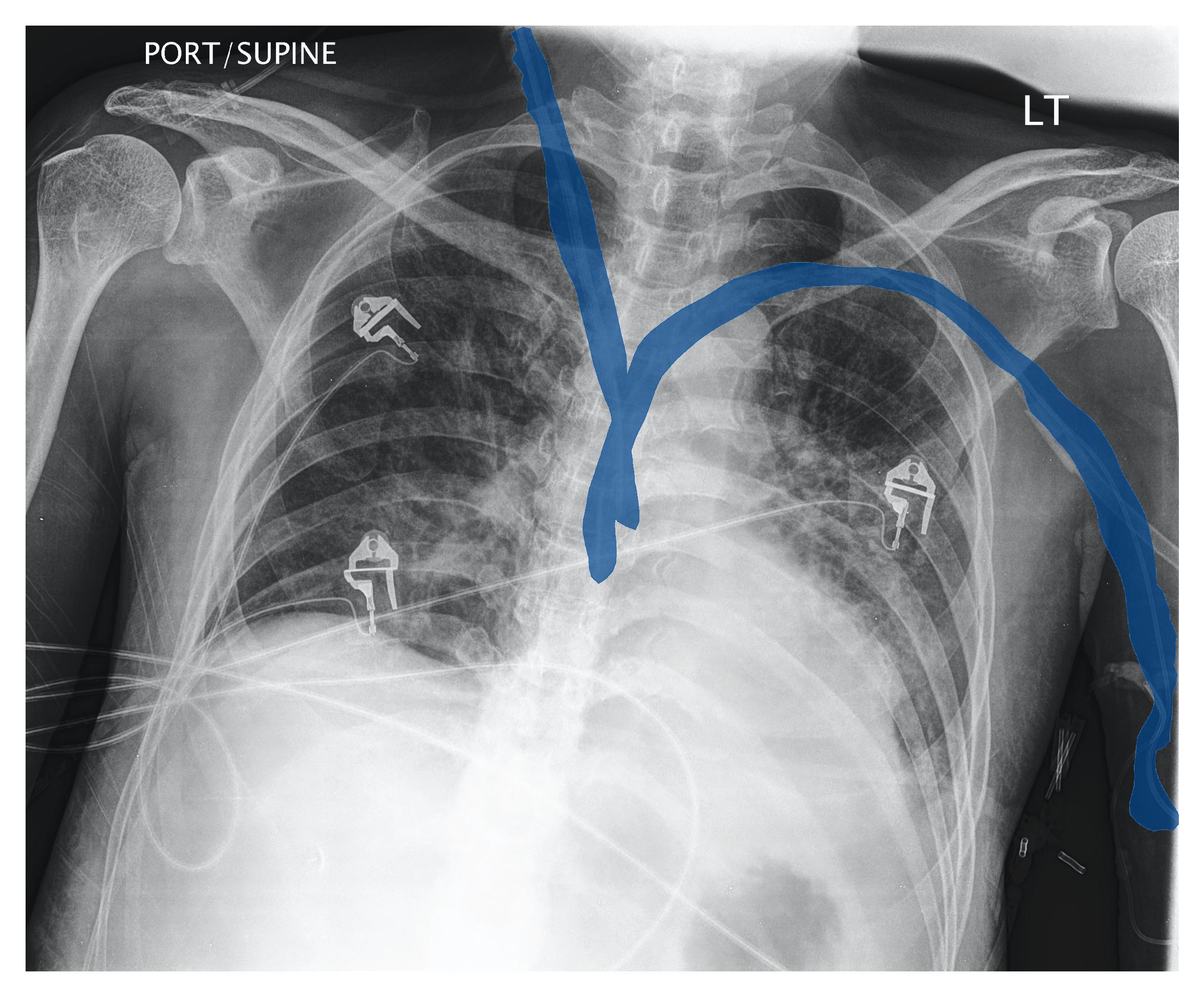}
        \caption{\textit{Support Devices} }
        \label{gt_multi_label_prob_Support_Devices}
    \end{subfigure}

    \caption{Multi-label instance of the \textit{Lung Opacity} (prediction probability score: 0.58) and \textit{Support Devices} (prediction probability score: 0.92) classes with human expert segmentation annotation for subsequent qualitative analysis.}
    
    \label{gt_multi_label_prob}
\end{figure}

In this qualitative study, our research objective is to explore the qualitative aspects of interpretability in two prominent pathology classes, namely \textit{Lung Opacity} and \textit{Support Devices}, which are widely utilized in the field. We visualize pixel-level segmentations of these selected pathology classes. We focused on these classes due to their high prediction performance, medical relevance, and prevalence within the domain. By employing various local interpretable methods, we aim to gain insights into how these methods shed light on the distinctive characteristics of the selected pathology classes.

For our qualitative analysis, we have established two criteria:

\begin{enumerate}

\item Examining the instance with the highest quantitative interpretability performance within these two classes. Through this analysis, we seek to evaluate how the quantitative performance can be interpreted qualitatively and whether this superior quantitative score aligns with the medical significance of the findings.

\item Additionally, we select the best predicted instance based on the prediction probability score for each of these classes, along with one instance exhibiting multiple labels. This additional selection allows us to investigate how different interpretability methods interpret the same instance in varying ways. Figure \ref{gt_single_prob_Lung_Opacity} depicts the chosen instance from the \textit{Lung Opacity} class, while Figure \ref{gt_single_prob_Support_Devices} showcases the instance from the \textit{Support Devices} class, both of which have been annotated with human expert segmentation. The accompanying figures also present the corresponding prediction probability scores. Figure \ref{gt_multi_label_prob} depicts the chosen multi-label instance for the \textit{Lung Opacity} and \textit{Support Devices} classes.

\end{enumerate}

\subsection{Medical Significance of the Selected Classes}\label{Medical_Significance}

\textit{Lung Opacity}, also referred to as "pulmonary opacification" \cite{hansell2008fleischner}, represents a common observation in chest radiography images. It signifies a decreased ratio of gas to soft tissue in the lungs. In a healthy lung, the appearance on a chest radiograph is black, indicating unobstructed airspaces. The presence of a gray or hazy region indicates the filling of airspaces with fluid, pus, or other substances. It may also indicate anomalies in the alveolar walls or thickening of the interstitial space between the lungs \cite{goodman2014felson}. \textit{Lung Opacity} can be a symptom of various diseases, including COVID-19 and pneumonia \cite{parekh2020review}.

\textit{Support Devices} refers to the identification of medical device(s) within the thoracic region. Prompt and accurate localization of these devices is crucial, as any potential complications need to be identified and addressed promptly to ensure patient safety. Various types of support devices exist, such as cardiac, endovascular, bronchopulmonary, and esophageal devices \cite{sigakis2018radiographic}.

\section{Results}

\begin{table}
\caption{Results of Classification on CheXpert Test Data}
\label{tab:chexpert_results}
\begin{tabular}{|l|l|l|}
\hline
\textbf{Pathology} & \textbf{AUROC} & \textbf{\begin{tabular}[c]{@{}l@{}}Positive\\ Count\end{tabular}} \\ \hline
No Finding & 0.79 & 38 \\ \hline
Cardiomegaly & 0.80 & 68 \\ \hline
Lung Lesion & 0.53 & 1 \\ \hline
Consolidation & 0.51 & 33 \\ \hline
Atelectasis & 0.53 & 80 \\ \hline
Pleural Effusion & 0.91 & 67 \\ \hline
Enlarged Cardiomediastinum & 0.60 & 109 \\ \hline
Lung Opacity & 0.87 & 126 \\ \hline
Edema & 0.87 & 45 \\ \hline
Pneumonia & 0.50 & 8 \\ \hline
Pneumothorax & 0.68 & 8 \\ \hline
Pleural Other & 1.00 & 1 \\ \hline
Support Devices & 0.89 & 107 \\ \hline
Fracture & - & 0 \\ \hline
\end{tabular}
\end{table}

\subsection{Pathology Prediction}

The evaluation of the model's performance is depicted in Table \ref{tab:chexpert_results}, illustrating the AUROC scores for each class. For this evaluation, a total of 234 input evaluation data points were utilized. However, it should be noted that the \textit{Fracture} class is absent in this evaluation due to the absence of any positive instances in the provided test data. The weighted average AUROC is determined to be 0.76. Additionally, the weighted average AUPRC is calculated to be 0.65.

An important observation is the considerable imbalance among positive instances across the classes. However, despite this imbalance, the results can be considered balanced. Notably, the \textit{Pleural Other} class exhibits noteworthy performance, even with only a single positive instance. These findings highlight the robustness and potential effectiveness of the model in handling varying class distributions and detecting positive instances accurately, even in challenging scenarios.

It is pertinent to note that the task of pathology prediction encompasses four additional classes in comparison to the quantitative evaluation tasks conducted using the CheXlocalize dataset. The CheXlocalize dataset employed in the quantitative evaluation comprised ten classes. The four supplementary classes introduced in the pathology prediction task include: \textit{No Finding}, \textit{Pleural Other}, \textit{Pneumonia}, and \textit{Fracture}. The inclusion of these classes expands the scope of the pathology prediction task and enables the identification and classification of a wider range of pathological conditions.

\subsection{LIME Interpretability Investigation}\label{lime_interpretability_evaluation}

\begin{figure*}[ht]
    \centering
    \includegraphics[width=.8\linewidth]{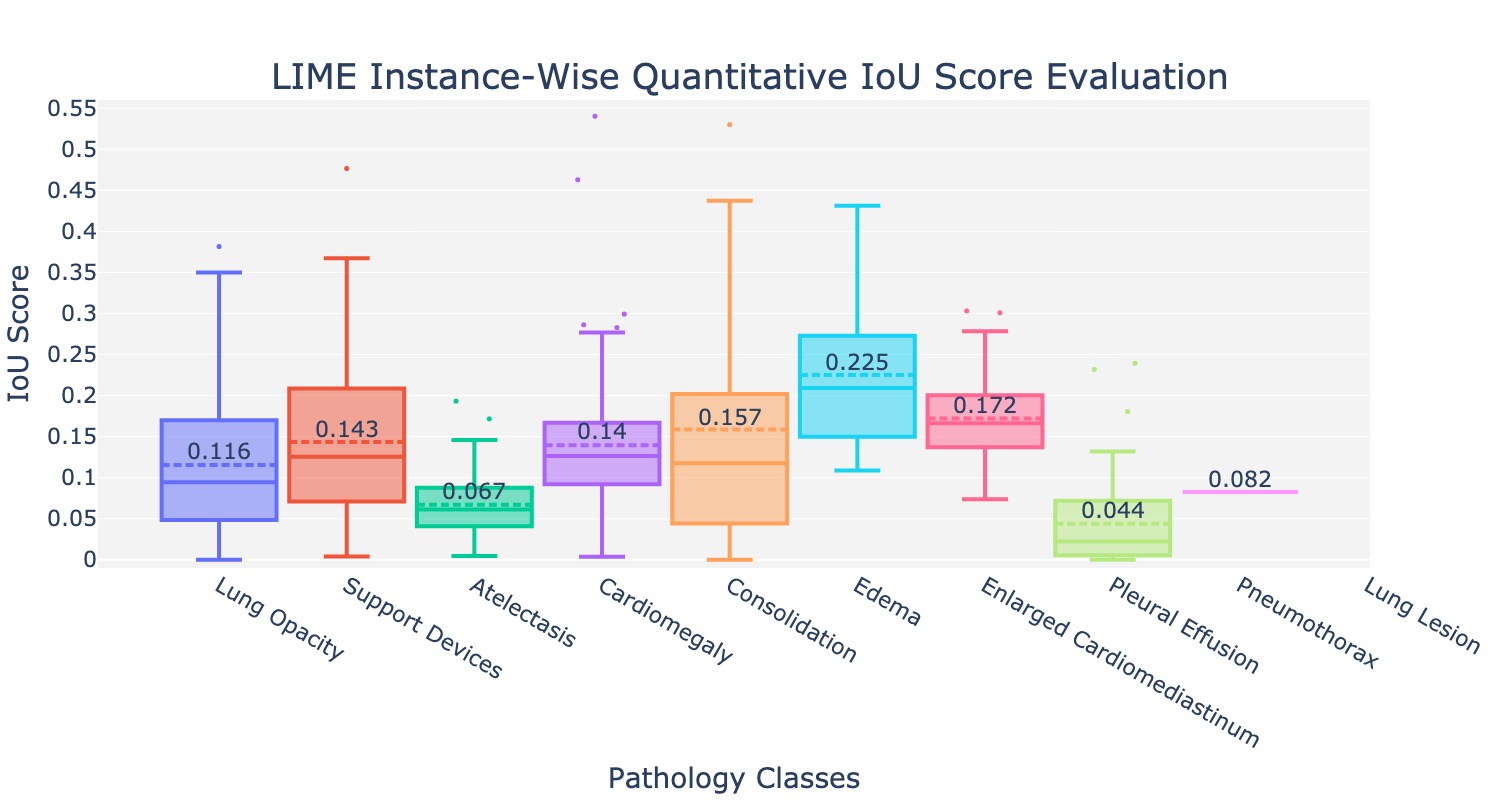}
    \caption{LIME instance-wise quantitative IoU score evaluation.}
    \label{lime_mean_iou}
\end{figure*}

Figure \ref{lime_mean_iou} depicts the evaluation of IoU scores, comparing the segmentation generated by the LIME interpretability method against human expert annotations. Each box-plot in the figure represents one of the ten pathology classes considered. The mean IoU values are indicated on each box-plot, while the dashed line represents the overall mean, and the solid line in close proximity to it signifies the median value. Notably, the results reveal that the LIME method yielded the most favorable outcomes for the \textit{Edema} class, exhibiting a mean score of 0.225. Conversely, it exhibited the poorest performance on the \textit{Lung Lesion} class, resulting in a score of zero. These findings provide valuable insights into the efficacy of the LIME interpretability method for different pathology classes.

\begin{figure}[ht]
    \centering
    \begin{subfigure}{.18\textwidth}
        \centering
        \includegraphics[width=1\linewidth]{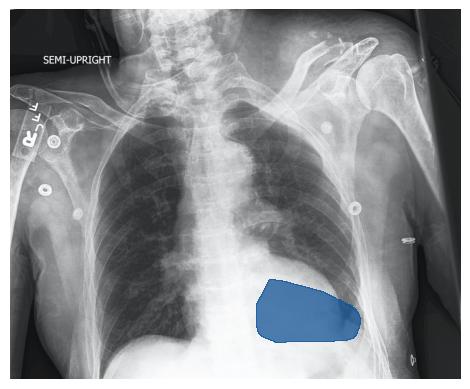}
        \caption{Ground Truth}
        \label{iou_gt_lime_Lung_Opacity}
    \end{subfigure} 
    \begin{subfigure}{.18\textwidth}
        \centering
        \includegraphics[width=1\linewidth]{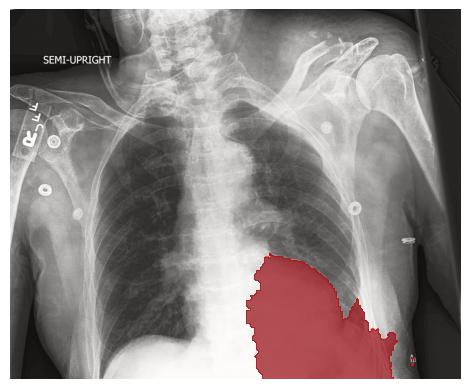}
        \caption{LIME Segment.}
        \label{iou_pred_lime_Lung_Opacity}
    \end{subfigure}

    \caption{LIME heatmap score segmentation of the \textit{Lung Opacity} (IoU score: 0.38, prediction probability score: 0.42) class with human expert segmentation annotation for comparison}
    
    \label{lime_iou_Lung_Opacity}
    
\end{figure}

Figure \ref{lime_iou_Lung_Opacity} presents the LIME segmentation visualization for the instance with the highest IoU score in the \textit{Lung Opacity} class. Upon examination, it is evident that LIME accurately identifies the lower-left lung region. However, there is a noticeable discrepancy between the LIME-generated segmentation and the human expert annotation, as LIME tends to encompass a larger area in its segmentation. Despite this slight disparity, it is noteworthy that LIME exhibits commendable performance in correctly identifying the relevant region in this instance. This observation underscores the effectiveness of the LIME method in localizing and delineating the target area for the \textit{Lung Opacity} class.

\begin{figure}[ht]
    \centering
    \begin{subfigure}{.18\textwidth}
        \centering
        \includegraphics[width=1\linewidth]{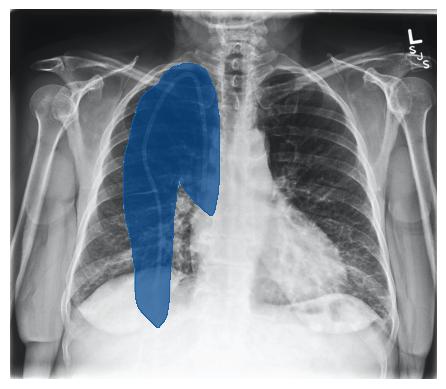}
        \caption{Ground Truth}
        \label{iou_gt_lime_Support_Devices}
    \end{subfigure} 
    \begin{subfigure}{.18\textwidth}
        \centering
        \includegraphics[width=1\linewidth]{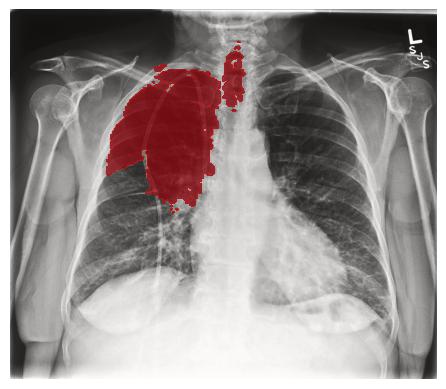}
        \caption{LIME Segment.}
        \label{iou_pred_lime_Support_Devices}
    \end{subfigure}

    \caption{LIME heatmap score segmentation of the best IoU scoring \textit{Support Devices} (IoU score: 0.48, prediction probability score: 0.86) class instance with human expert segmentation annotation for comparison.}
    
    \label{lime_iou_Support_Devices}
\end{figure}

Figure \ref{lime_iou_Support_Devices} displays the LIME segmentation visualization for the instance with the highest IoU score in the \textit{Support Devices} class. Upon examination, it is evident that LIME accurately identifies the upper-right lung region, successfully localizing the target area of interest. However, it is important to note that the LIME-generated segmentation does not entirely encompass the entirety of the corresponding human expert annotation area. Despite this partial coverage, the result demonstrates LIME's ability to effectively identify and localize the upper-right lung region for instances within the \textit{Support Devices} class. This finding underscores the utility of the LIME interpretability method in identifying regions of interest, although further refinement may be necessary to achieve full correspondence with human expert annotations.

\begin{figure}[ht]
    \centering
    \begin{subfigure}{.18\textwidth}
        \centering
        \includegraphics[width=1\linewidth]{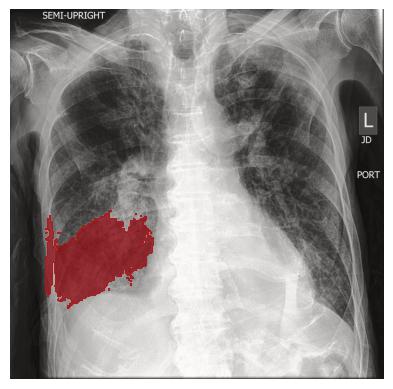}
        \caption{\textit{Lung Opacity}}
        \label{lime_pred_Lung_Opacity_single_prob}
    \end{subfigure} 
    \begin{subfigure}{.18\textwidth}
        \centering
        \includegraphics[width=1\linewidth]{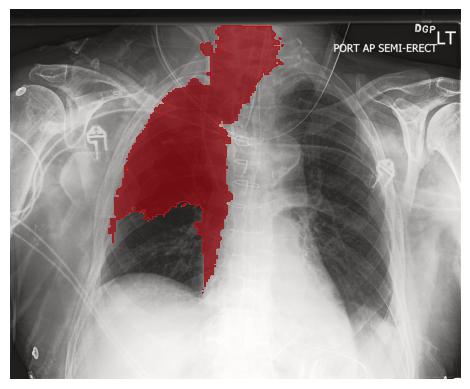}
        \caption{\textit{Support Devices}}
        \label{lime_pred_Support_Devices_single_prob}
    \end{subfigure}

    \caption{LIME heatmap score segmentation of the best prediction probability scoring \textit{Lung Opacity} (prediction probability score: 0.63, IoU score: 0.16) and \textit{Support Devices} (prediction probability score: 0.93, IoU score: 0.25) class instances.}
    
    \label{lime_single_prob}
\end{figure}

Figure \ref{lime_single_prob} showcases the LIME segmentation visualization for the instance with the highest prediction probability in both the \textit{Lung Opacity} and \textit{Support Devices} classes. A comparative analysis with Figure \ref{gt_single_prob}, which presents the human expert annotation segmentation for these classes, allows for insightful observations. In the case of the \textit{Lung Opacity} class, the LIME segmentation only partially covers one of the three designated areas. Conversely, in the \textit{Support Devices} class, the LIME segmentation predominantly focuses on the upper-right lung region, overlapping with and partially covering the corresponding human expert annotation segmentation area, particularly its left section.

\begin{figure}[ht]
    \centering
    \begin{subfigure}{.18\textwidth}
        \centering
        \includegraphics[width=1\linewidth]{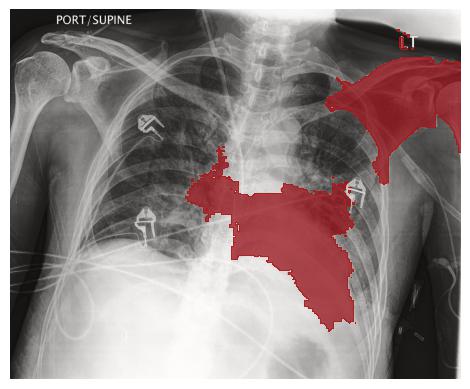}
        \caption{\textit{Lung Opacity}}
        \label{lime_pred_Lung_Opacity_multi_label}
    \end{subfigure} 
    \begin{subfigure}{.18\textwidth}
        \centering
        \includegraphics[width=1\linewidth]{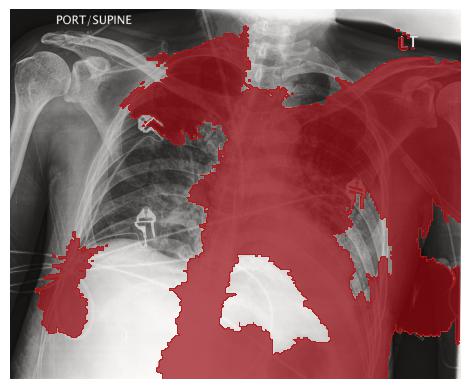}
        \caption{\textit{Support Devices}}
        \label{lime_pred_Support_Devices_multi_label}
    \end{subfigure}

    \caption{LIME heatmap score segmentation of the best prediction probability scoring multi-label class instance (\textit{Lung Opacity} - prediction probability score: 0.58, IoU score: 0.31 and \textit{Support Devices} - prediction probability score: 0.92, IoU score: 0.08).}
    
    \label{lime_multi_label}
\end{figure}

Figure \ref{lime_multi_label} depicts the LIME segmentation visualization for the instance with the highest prediction probability among the multi-label instances in both the \textit{Lung Opacity} and \textit{Support Devices} classes. A comparison with Figure \ref{gt_multi_label_prob}, which presents the corresponding human expert annotation segmentation for these classes, allows for insightful analysis.

In the case of the \textit{Lung Opacity} class, the LIME segmentation partially identifies two of the three designated segmentation regions, aligning with certain areas of interest. However, it is important to note that LIME mistakenly identifies the left shoulder and an artifact situated on top of it as relevant regions, despite their lack of association with the \textit{Lung Opacity} class. This highlights a potential limitation in the LIME method's ability to accurately discern between the true areas of interest and extraneous artifacts.

Turning to the \textit{Support Devices} class, the LIME segmentation covers a substantial portion of the left side of the body, incorrectly marking both the relevant region and an artifact positioned on top of it. This indicates a lack of precision in the LIME method's identification and localization of the target region within the \textit{Support Devices} class.

These observations emphasize the need for further refinement of the LIME segmentation approach to ensure accurate and reliable identification of the relevant regions while minimizing misclassifications of artifacts and unrelated areas. Further refinement is warranted to improve its consistency and alignment with human expert annotation segmentations for a more robust and reliable analysis.

\subsection{SHAP Interpretability Investigation}\label{shap_interpretability_evaluation}

\begin{figure*}[ht]
    \centering
    \includegraphics[width=.8\linewidth]{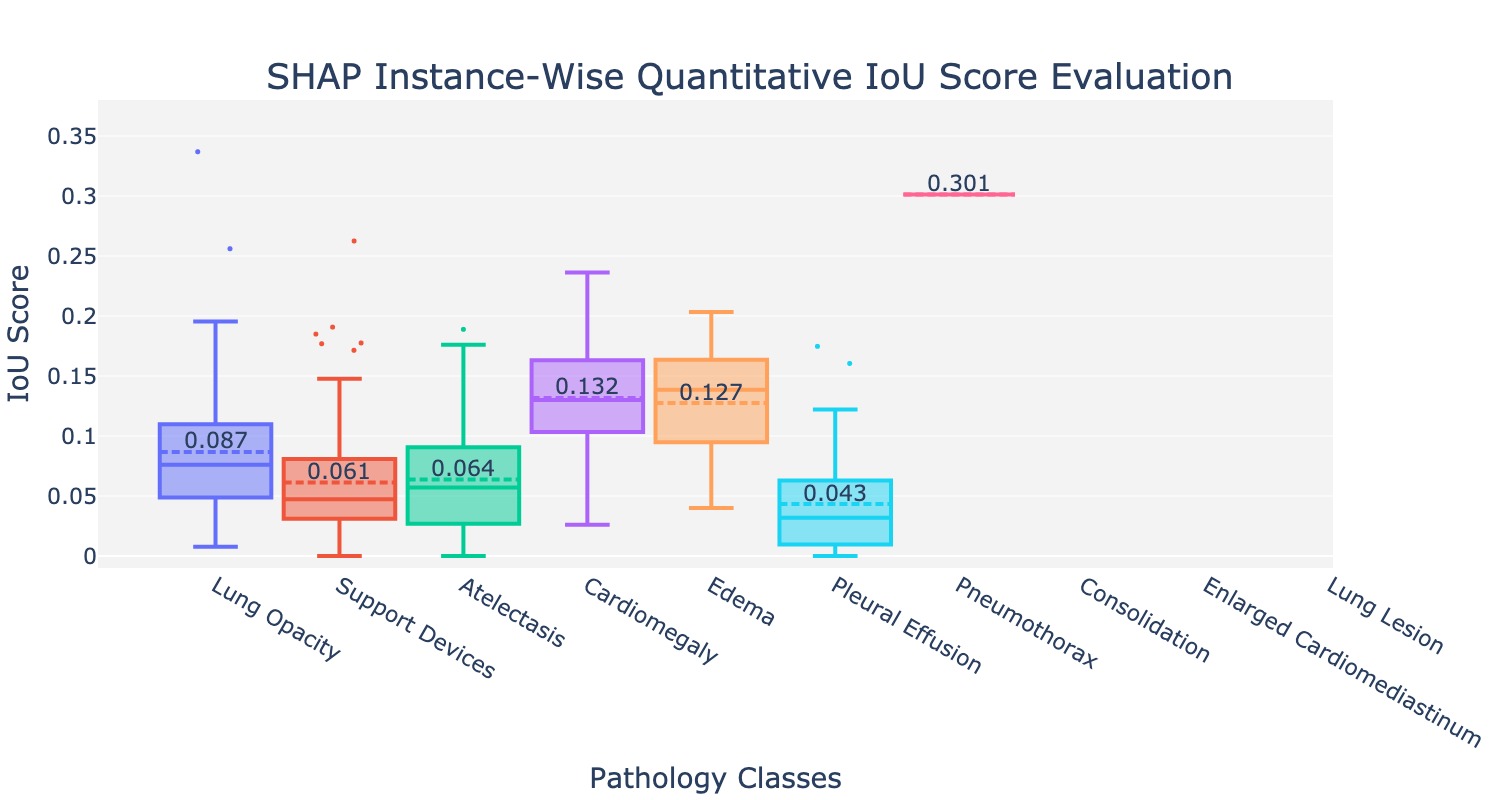}
    \caption{SHAP instance-wise quantitative IoU score evaluation.}
    \label{shap_mean_iou}
\end{figure*}

Figure \ref{shap_mean_iou} presents the IoU evaluation scores, comparing the SHAP interpretability method's generated segmentation against the corresponding human expert annotation. An examination of the results reveals that the SHAP method did not demonstrate robust generalization across all pathology classes. Particularly, for three classes, namely \textit{Consolidation}, \textit{Enlarged Cardiomediastinum}, and \textit{Lung Lesion}, SHAP failed to provide any meaningful values for any of the instances assessed. This indicates a lack of effectiveness in accurately identifying and segmenting these specific pathology classes using the SHAP method.

Furthermore, the highest mean IoU score of 0.301 was achieved for the \textit{Pneumothorax} class. In contrast, the mean IoU scores for the remaining classes were considerably lower when compared to this top score, indicating comparatively poorer performance in accurately capturing the relevant regions for those classes using the SHAP method.

\begin{figure}[ht]
    \centering
    \begin{subfigure}{.18\textwidth}
        \centering
        \includegraphics[width=1\linewidth]{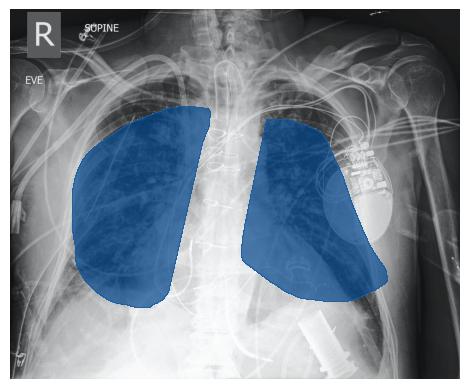}
        \caption{Ground Truth}
        \label{iou_gt_shap_Lung_Opacity}
    \end{subfigure} 
    \begin{subfigure}{.18\textwidth}
        \centering
        \includegraphics[width=1\linewidth]{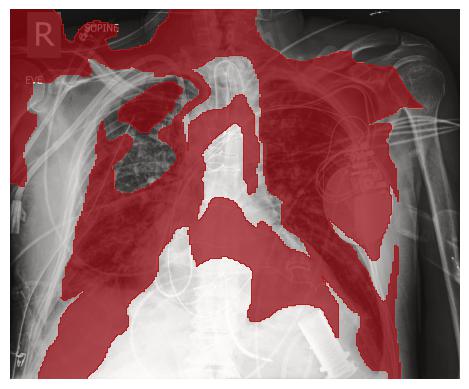}
        \caption{SHAP Segment.}
        \label{iou_pred_shap_Lung_Opacity}
    \end{subfigure}

    \caption{SHAP heatmap score segmentation of the \textit{Lung Opacity} (IoU score: 0.34, prediction probability score: 0.52) class with human expert segmentation annotation for comparison.}
    
    \label{shap_iou_Lung_Opacity}
\end{figure}

Figure \ref{shap_iou_Lung_Opacity} displays the SHAP segmentation visualization for the instance with the highest IoU score within the \textit{Lung Opacity} class. A careful examination of the visualization reveals that the SHAP segmentation successfully covers the region annotated by human experts, effectively localizing the target area of interest. However, it is important to note that the SHAP segmentation also erroneously encompasses both shoulders, the neck area, and an artifact located in the top-right area.

\begin{figure}[ht]
    \centering
    \begin{subfigure}{.18\textwidth}
        \centering
        \includegraphics[width=1\linewidth]{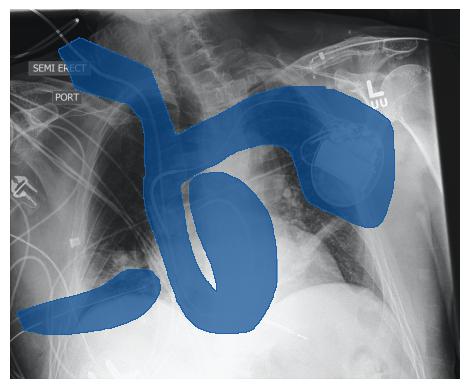}
        \caption{Ground Truth}
        \label{iou_gt_shap_Support_Devices}
    \end{subfigure} 
    \begin{subfigure}{.18\textwidth}
        \centering
        \includegraphics[width=1\linewidth]{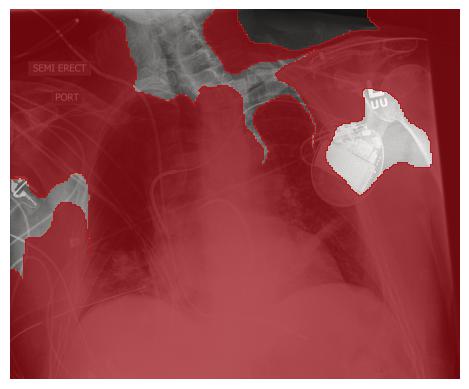}
        \caption{SHAP Segment.}
        \label{iou_pred_shap_Support_Devices}
    \end{subfigure}

    \caption{SHAP heatmap score segmentation of the best IoU scoring \textit{Support Devices} (IoU score: 0.26, prediction probability score: 0.63) class instance with human expert segmentation annotation for comparison.}
    
    \label{shap_iou_Support_Devices}
\end{figure}

Figure \ref{shap_iou_Support_Devices} illustrates the SHAP segmentation visualization for the instance with the highest IoU score within the \textit{Support Devices} class. The SHAP segmentation displays notable overlap with the ground truth areas, successfully capturing certain regions of interest. However, it is important to note that the SHAP segmentation is extensive and entirely overlaps the bottom region of the image, thereby leading to incorrect labeling.

This observation highlights a limitation in the SHAP segmentation's ability to accurately delineate and localize the relevant region within the \textit{Support Devices} class. Although it achieves a reasonable degree of correspondence with the ground truth areas, the excessive coverage beyond the intended region compromises its accuracy and specificity.

\begin{figure}[ht]
    \centering
    \begin{subfigure}{.18\textwidth}
        \centering
        \includegraphics[width=1\linewidth]{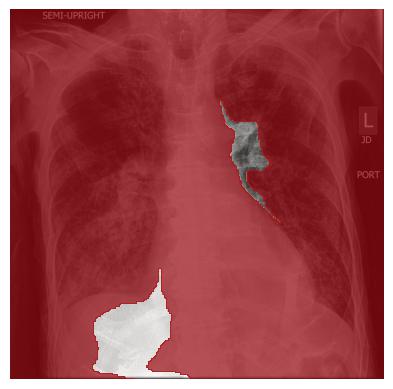}
        \caption{\textit{Lung Opacity}}
        \label{shap_pred_Lung_Opacity_single_prob}
    \end{subfigure} 
    \begin{subfigure}{.18\textwidth}
        \centering
        \includegraphics[width=1\linewidth]{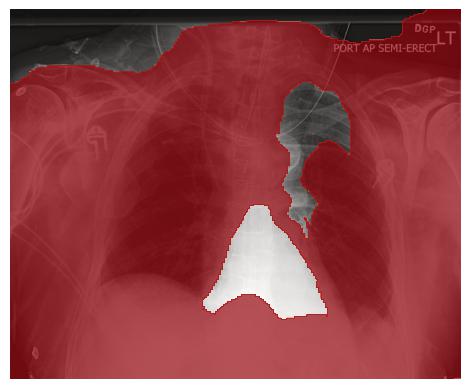}
        \caption{\textit{Support Devices}}
        \label{shap_pred_Support_Devices_single_prob}
    \end{subfigure}

    \caption{SHAP heatmap score segmentation of the best prediction probability scoring \textit{Lung Opacity} (prediction probability score: 0.63, IoU score: 0.23) and \textit{Support Devices} (prediction probability score: 0.93, IoU score: 0.13) class instances.}
    
    \label{shap_single_prob}
\end{figure}

Figure \ref{shap_single_prob} displays the SHAP segmentation visualization for the instance with the highest prediction probability in both the \textit{Lung Opacity} and \textit{Support Devices} classes. A comparison with Figure \ref{gt_single_prob}, which presents the corresponding human expert annotation segmentation for these classes, allows for a comprehensive analysis.

Upon examination, it becomes apparent that the SHAP segmentation almost entirely covers the entire image for both classes. While this extensive coverage includes the correct regions of interest, the overwhelming extent of the segmentation diminishes its interpretive significance. Consequently, the SHAP segmentation, despite accurately capturing the relevant regions, lacks specificity and discriminative power due to its indiscriminate and excessive nature.

\begin{figure}[ht]
    \centering
    \begin{subfigure}{.18\textwidth}
        \centering
        \includegraphics[width=1\linewidth]{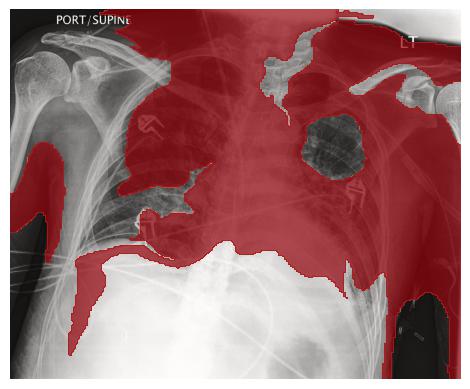}
        \caption{\textit{Lung Opacity}}
        \label{shap_pred_Lung_Opacity_multi_label}
    \end{subfigure} 
    \begin{subfigure}{.18\textwidth}
        \centering
        \includegraphics[width=1\linewidth]{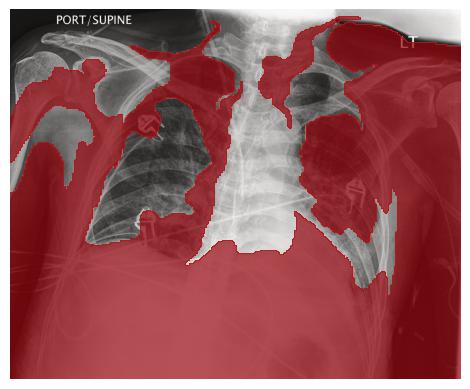}
        \caption{\textit{Support Devices}}
        \label{shap_pred_Support_Devices_multi_label}
    \end{subfigure}

    \caption{SHAP heatmap score segmentation of the best prediction probability scoring multi-label class instance (\textit{Lung Opacity} - prediction probability score: 0.58, IoU score: 0.17 and \textit{Support Devices} - prediction probability score: 0.92, IoU score: 0.03).}
    
    \label{shap_multi_label}
\end{figure}

Figure \ref{shap_multi_label} presents the SHAP segmentation visualization for the instance with the highest prediction probability among the multi-label instances in both the \textit{Lung Opacity} and \textit{Support Devices} classes. A comparative analysis with Figure \ref{gt_multi_label_prob}, which displays the human expert annotation segmentation for these classes, enables a comprehensive examination of the results.

In terms of the \textit{Lung Opacity} class, the SHAP segmentation demonstrates partial identification of two out of the three designated segmentation regions. However, it incorrectly identifies the left shoulder, arm, neck, and an artifact located in the left-top area of the body as relevant regions. Moreover, the SHAP segmentation exhibits a somewhat extensive nature in comparison to the ground truth segmentation regions. Consequently, while the SHAP segmentation captures certain areas of interest, its inclusion of unrelated structures and artifacts diminishes its overall accuracy and interpretive value.

Regarding the \textit{Support Devices} class, the SHAP segmentation covers the entire regions of the image except for the middle part. However, when compared to Figure \ref{gt_multi_label_prob_Support_Devices}, the relevance of the SHAP segmentation is insignificant. The segmentation fails to accurately capture and delineate the specific areas of interest within the \textit{Support Devices} class, highlighting a limitation in the SHAP method's ability to provide precise and discriminative segmentations for this class.

These observations underscore a limitation in the SHAP interpretability method's ability to provide concise and focused segmentations. Although it successfully identifies the areas of interest, the lack of precision and the encompassing nature of the SHAP segmentation limit its interpretive value.

\subsection{Grad-CAM Interpretability Investigation}\label{grad_cam_interpretability_evaluation}

\begin{figure*}[ht]
    \centering
    \includegraphics[width=.8\linewidth]{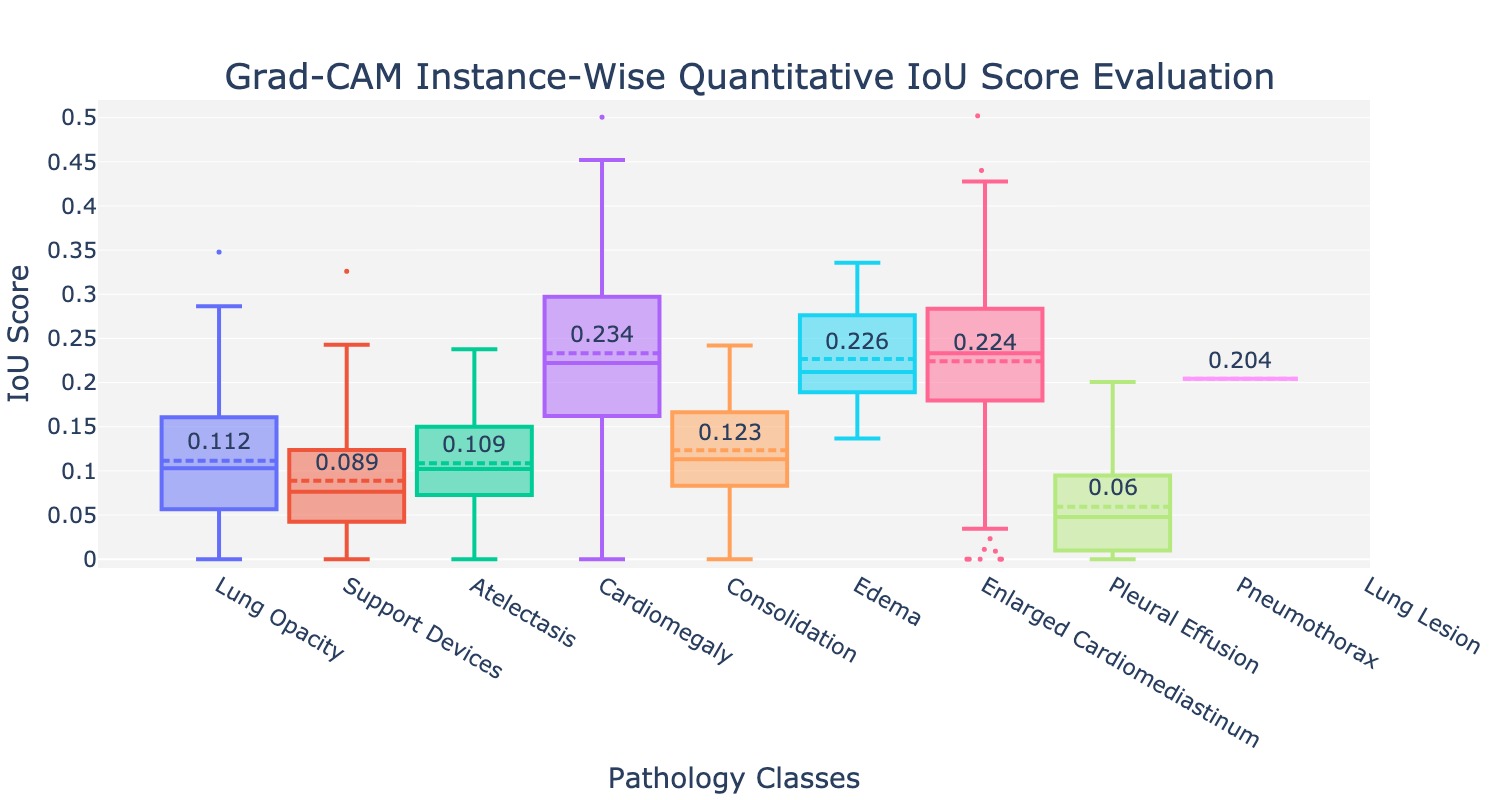}
    \caption{Grad-CAM instance-wise quantitative IoU score evaluation.}
    \label{grad_cam_mean_iou}
\end{figure*}

Figure \ref{grad_cam_mean_iou} presents the IoU evaluation scores, comparing the Grad-CAM interpretability method's generated segmentation against the corresponding human expert annotation. The results demonstrate a balanced performance of Grad-CAM across all evaluated pathology classes.

Notably, the \textit{Cardiomegaly} class achieves the highest mean IoU score of 0.234. Conversely, the \textit{Lung Lesion} class attains the worst mean IoU score of zero. This score signifies a lack of correspondence and agreement between the Grad-CAM segmentation and the human expert annotation for instances within this class.

\begin{figure}[ht]
    \centering
    \begin{subfigure}{.18\textwidth}
        \centering
        \includegraphics[width=1\linewidth]{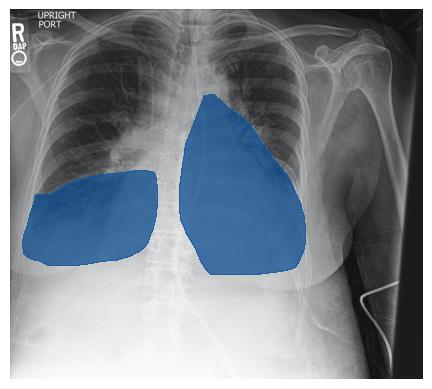}
        \caption{Ground Truth}
        \label{iou_gt_grad_cam_Lung_Opacity}
    \end{subfigure} 
    \begin{subfigure}{.18\textwidth}
        \centering
        \includegraphics[width=1\linewidth]{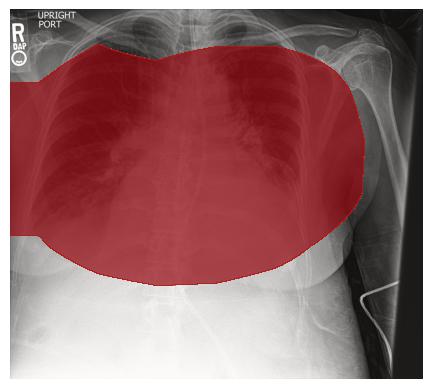}
        \caption{Grad-CAM Seg.}
        \label{iou_pred_grad_cam_Lung_Opacity}
    \end{subfigure}

    \caption{Grad-CAM heatmap score segmentation of the \textit{Lung Opacity} (IoU score: 0.35, prediction probability score: 0.54) class with human expert segmentation annotation for comparison.}
    
    \label{grad_cam_iou_Lung_Opacity}
\end{figure}

Figure \ref{grad_cam_iou_Lung_Opacity} presents the Grad-CAM segmentation visualization for the instance with the highest IoU score within the \textit{Lung Opacity} class. Upon examination, it becomes apparent that the Grad-CAM segmentation successfully encompasses the region annotated by human experts, accurately localizing the intended area of interest. However, it is important to note that the Grad-CAM segmentation exhibits a broad and expansive nature, which can hinder the ability to distinguish the precise boundaries of the human expert annotated region.

\begin{figure}[ht]
    \centering
    \begin{subfigure}{.18\textwidth}
        \centering
        \includegraphics[width=1\linewidth]{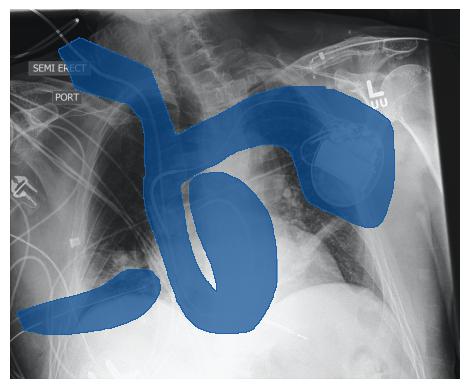}
        \caption{Ground Truth}
        \label{iou_gt_grad_cam_Support_Devices}
    \end{subfigure} 
    \begin{subfigure}{.18\textwidth}
        \centering
        \includegraphics[width=1\linewidth]{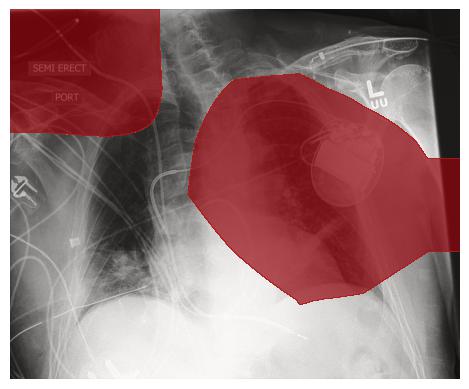}
        \caption{Grad-CAM Seg.}
        \label{iou_pred_grad_cam_Support_Devices}
    \end{subfigure}

    \caption{Grad-CAM heatmap score segmentation of the best IoU scoring \textit{Support Devices} (IoU score: 0.32, prediction probability score: 0.63) class instance with human expert segmentation annotation for comparison.}
    
    \label{grad_cam_iou_Support_Devices}
\end{figure}

Figure \ref{grad_cam_iou_Support_Devices} illustrates the Grad-CAM segmentation visualization for the instance with the highest IoU score within the \textit{Support Devices} class. Upon observation, it is evident that the Grad-CAM segmentation successfully captures certain regions in the top-right and middle-left parts of the body that correspond to the correct areas of interest within the \textit{Support Devices} class. However, similar to previous cases, the Grad-CAM segmentation exhibits a relatively large coverage area compared to the human expert annotated regions.

\begin{figure}[ht]
    \centering
    \begin{subfigure}{.18\textwidth}
        \centering
        \includegraphics[width=1\linewidth]{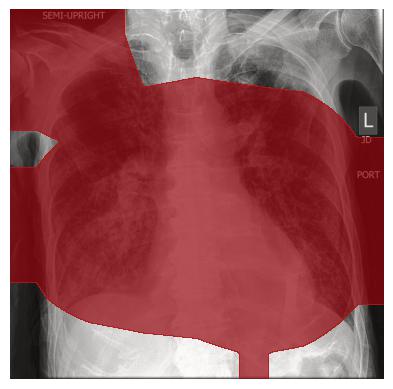}
        \caption{\textit{Lung Opacity}}
        \label{grad_cam_pred_Lung_Opacity_single_prob}
    \end{subfigure} 
    \begin{subfigure}{.18\textwidth}
        \centering
        \includegraphics[width=1\linewidth]{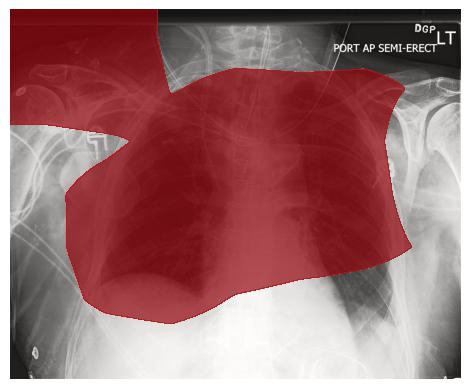}
        \caption{\textit{Support Devices}}
        \label{grad_cam_pred_Support_Devices_single_prob}
    \end{subfigure}

    \caption{Grad-CAM heatmap score segmentation of the best prediction probability scoring \textit{Lung Opacity} (prediction probability score: 0.63, IoU score: 0.23) and \textit{Support Devices} (prediction probability score: 0.93, IoU score: 0.18) class instances.}
    
    \label{grad_cam_single_prob}
\end{figure}

Figure \ref{grad_cam_single_prob} exhibits the Grad-CAM segmentation visualization for the instance with the highest prediction probability in both the \textit{Lung Opacity} and \textit{Support Devices} classes. A comparison with Figure \ref{gt_single_prob}, which displays the corresponding human expert annotation segmentation for these classes, enables a comprehensive analysis.

Upon examination, it becomes evident that the Grad-CAM segmentation, while capturing the relevant regions of interest, covers a significantly broader area compared to the human expert annotation segmentations for both classes. This expansive coverage, while encompassing the correct regions, poses challenges in precisely identifying the precise boundaries of the relevant regions. Consequently, the Grad-CAM segmentations may not offer the level of nuance required for a nuanced understanding of the medical significance of the identified regions.

\begin{figure}[ht]
    \centering
    \begin{subfigure}{.18\textwidth}
        \centering
        \includegraphics[width=1\linewidth]{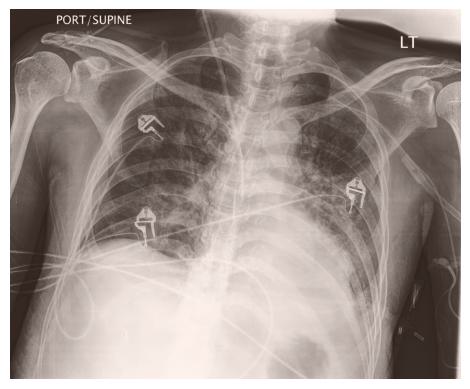}
        \caption{\textit{Lung Opacity}}
        \label{grad_cam_pred_Lung_Opacity_multi_label}
    \end{subfigure} 
    \begin{subfigure}{.18\textwidth}
        \centering
        \includegraphics[width=1\linewidth]{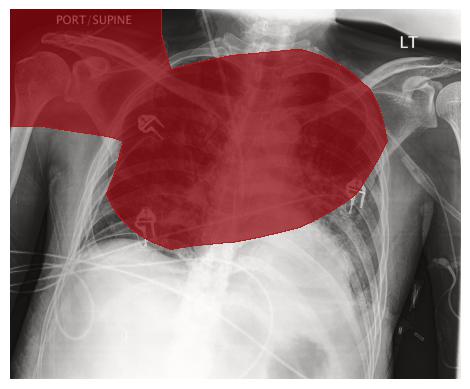}
        \caption{\textit{Support Devices}}
        \label{grad_cam_pred_Support_Devices_multi_label}
    \end{subfigure}

    \caption{Grad-CAM heatmap score segmentation of the best prediction probability scoring multi-label class instance (\textit{Lung Opacity} - prediction probability score: 0.58, IoU score: NA and \textit{Support Devices} - prediction probability score: 0.92, IoU score: 0.06).}
    
    \label{grad_cam_multi_label}
\end{figure}

In terms of the \textit{Lung Opacity} class, the Grad-CAM segmentation fails to identify any relevant area within the image. Turning to the \textit{Support Devices} class, the Grad-CAM segmentation exhibits extensive coverage of the middle and upper-right area of the body. However, when compared to Figure \ref{gt_multi_label_prob_Support_Devices}, the relevance of the Grad-CAM segmentation becomes insignificant. The broad coverage of non-relevant areas diminishes the interpretability and discriminative power of the Grad-CAM method for accurately identifying the specific regions of interest within the \textit{Support Devices} class.

Upon examination, it becomes evident that the Grad-CAM segmentation, while capturing the relevant regions of interest, covers a significantly broader area compared to the human expert annotation segmentations for both classes. This expansive coverage, while encompassing the correct regions, poses challenges in precisely identifying the precise boundaries of the relevant regions. Consequently, the Grad-CAM segmentations may not offer the level of nuance required for a nuanced understanding of the medical significance of the identified regions.

\subsection{LRP Interpretability Investigation}\label{lrp_interpretability_evaluation}

\begin{figure*}[ht]
    \centering
    \includegraphics[width=.8\linewidth]{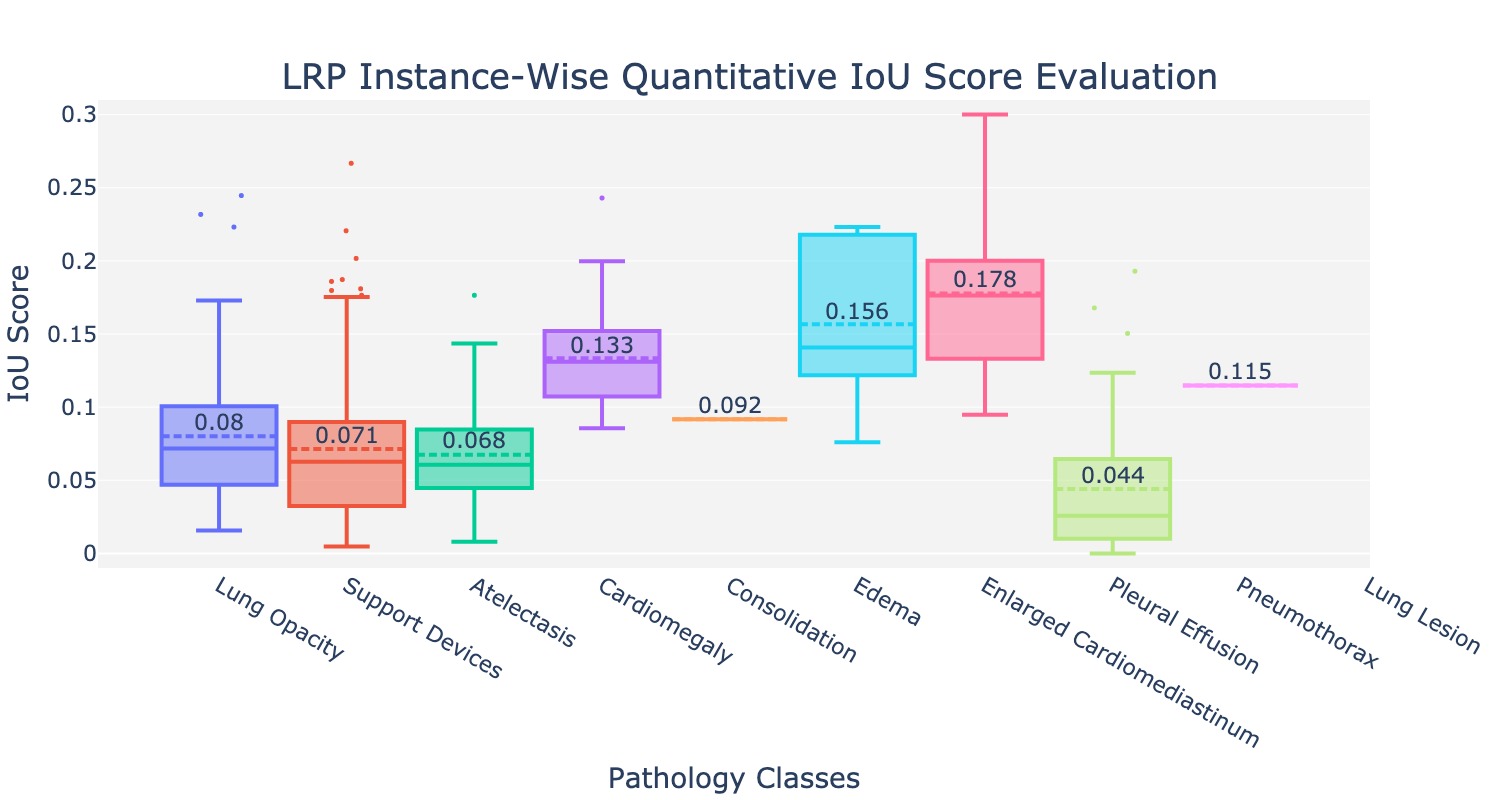}
    \caption{LRP instance-wise quantitative IoU score evaluation.}
    \label{lrp_mean_iou}
\end{figure*}

Figure \ref{lrp_mean_iou} presents the IoU evaluation scores, which compare the LRP interpretability method's generated segmentation against the corresponding human expert annotation. The results reveal valuable insights into the performance of the LRP method in accurately delineating the regions of interest.

Among the evaluated classes, the \textit{Enlarged Cardiomediastinum} class achieved the highest mean IoU score, with a value of 0.178. This indicates a relatively better alignment between the LRP-generated segmentation and the human expert annotation for this particular class. The higher mean IoU score suggests a higher degree of correspondence and accuracy in identifying and delineating the relevant regions. On the other hand, the \textit{Lung Lesion} class obtained the worst mean IoU score, with a value of zero. 

\begin{figure}[ht]
    \centering
    \begin{subfigure}{.18\textwidth}
        \centering
        \includegraphics[width=1\linewidth]{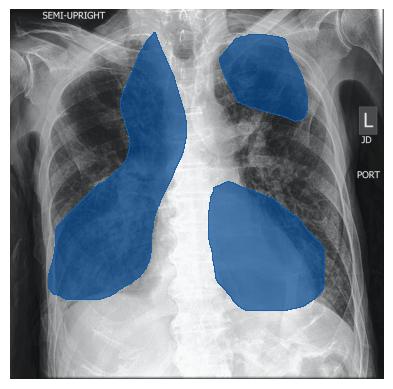}
        \caption{Ground Truth}
        \label{iou_gt_lrp_Lung_Opacity}
    \end{subfigure} 
    \begin{subfigure}{.18\textwidth}
        \centering
        \includegraphics[width=1\linewidth]{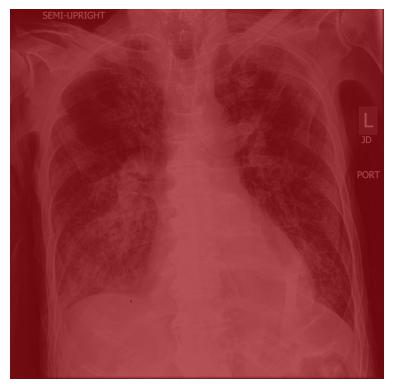}
        \caption{LRP Segment.}
        \label{iou_pred_lrp_Lung_Opacity}
    \end{subfigure}

    \caption{LRP heatmap score segmentation of the \textit{Lung Opacity} (IoU score: 0.24, prediction probability score: 0.63) class with human expert segmentation annotation for comparison.}
    
    \label{lrp_iou_Lung_Opacity}
\end{figure}

\begin{figure}[ht]
    \centering
    \begin{subfigure}{.18\textwidth}
        \centering
        \includegraphics[width=1\linewidth]{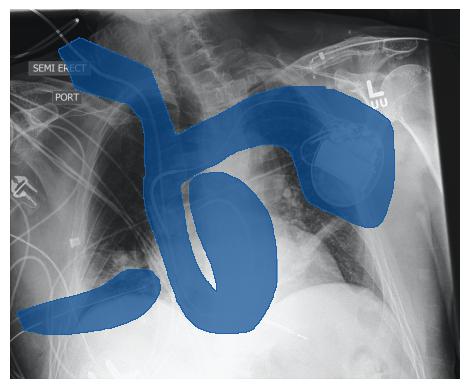}
        \caption{Ground Truth}
        \label{iou_gt_lrp_Support_Devices}
    \end{subfigure} 
    \begin{subfigure}{.18\textwidth}
        \centering
        \includegraphics[width=1\linewidth]{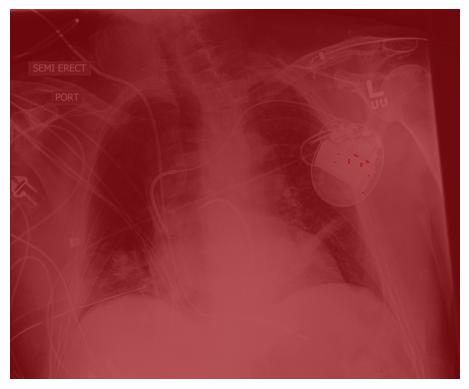}
        \caption{LRP Segment.}
        \label{iou_pred_lrp_Support_Devices}
    \end{subfigure}

    \caption{LRP heatmap score segmentation of the best IoU scoring \textit{Support Devices} (IoU score: 0.27, prediction probability score: 0.63) class instance with human expert segmentation annotation for comparison.}
    
    \label{lrp_iou_Support_Devices}
\end{figure}

\begin{figure}[ht]
    \centering
    \begin{subfigure}{.18\textwidth}
        \centering
        \includegraphics[width=1\linewidth]{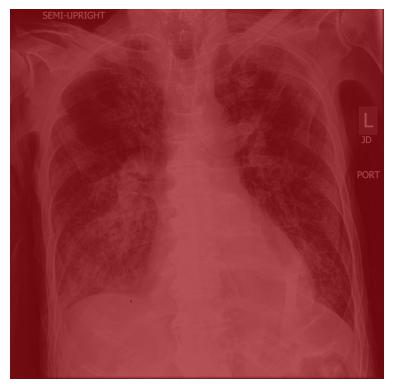}
        \caption{\textit{Lung Opacity}}
        \label{lrp_pred_Lung_Opacity_single_prob}
    \end{subfigure} 
    \begin{subfigure}{.18\textwidth}
        \centering
        \includegraphics[width=1\linewidth]{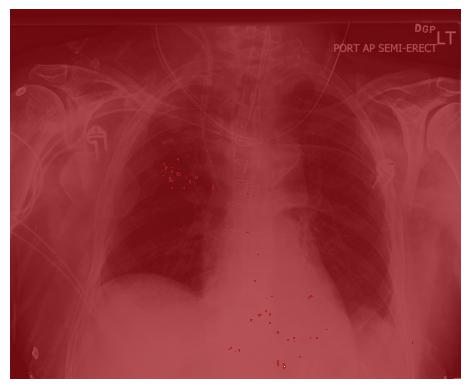}
        \caption{\textit{Support Devices}}
        \label{lrp_pred_Support_Devices_single_prob}
    \end{subfigure}

    \caption{LRP heatmap score segmentation of the best prediction probability scoring \textit{Lung Opacity} (prediction probability score: 0.63, IoU score: 0.24) and \textit{Support Devices} (prediction probability score: 0.93, IoU score: 0.14) class instances.}
    
    \label{lrp_single_prob}
\end{figure}

\begin{figure}[ht]
    \centering
    \begin{subfigure}{.18\textwidth}
        \centering
        \includegraphics[width=1\linewidth]{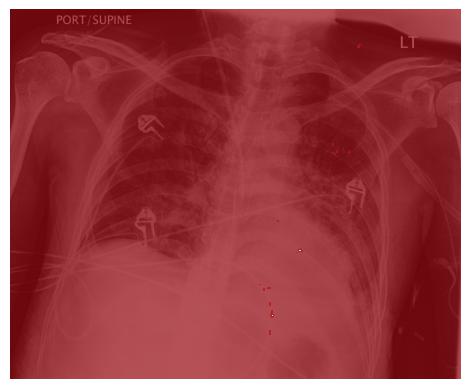}
        \caption{\textit{Lung Opacity}}
        \label{lrp_pred_Lung_Opacity_multi_label}
    \end{subfigure} 
    \begin{subfigure}{.18\textwidth}
        \centering
        \includegraphics[width=1\linewidth]{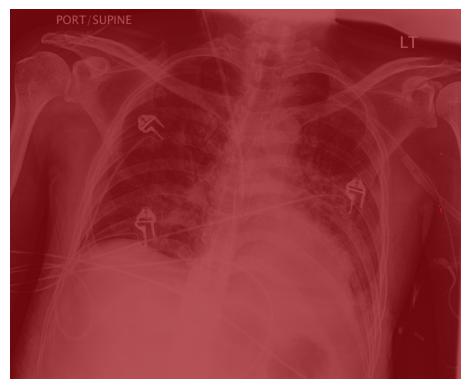}
        \caption{\textit{Support Devices}}
        \label{lrp_pred_Support_Devices_multi_label}
    \end{subfigure}

    \caption{LRP heatmap score segmentation of the best prediction probability scoring multi-label class instance (\textit{Lung Opacity} - prediction probability score: 0.58, IoU score: 0.12 and \textit{Support Devices} - prediction probability score: 0.92, IoU score: 0.04).}
    
    \label{lrp_multi_label}
\end{figure}

The qualitative visualizations of all the cases, as presented in Figure \ref{lrp_iou_Lung_Opacity}, \ref{lrp_iou_Support_Devices}, \ref{lrp_single_prob}, and \ref{lrp_multi_label}, reveal a consistent pattern. Across these images, the LRP segmentation covers the entirety of the image regions, exhibiting a uniform and indiscriminate nature. Consequently, the LRP heatmap score visualizations lack class discriminability, exhibit considerable noise, and fail to provide meaningful insights specific to any particular classes. While the quantitative evaluation in Figure \ref{lrp_mean_iou} may yield acceptable values, the qualitative analysis underscores the limited medical significance of the LRP method.

The visual examination of the LRP visualizations highlights the absence of focused and interpretable segmentations. The complete coverage of image regions by the LRP segmentation masks inhibits its ability to provide precise and discriminative interpretations. This lack of specificity diminishes the utility and usefulness of the LRP method from a medical significance standpoint.

Thus, while the quantitative evaluation may provide some acceptable metrics, the qualitative analysis reveals inherent limitations in the LRP interpretability method. The non-discriminative, noisy, and indiscriminate nature of the LRP heatmap score visualizations diminishes their practical value in medical image analysis. Future research endeavors should focus on refining the LRP method to address these limitations and develop more precise and medically significant visualizations.

\section{Discussion}
\label{discussion}

This study undertakes an exploration of four prominent local interpretability methods commonly utilized in the field, namely LIME, SHAP, Grad-CAM, and LRP. Our primary objective is to comprehensively examine the applicability of these methods through (1) a quantitative evaluation of the heatmap score performance across all pathology classes within a multi-label-multi-class pathology prediction task. Furthermore, we aim to (2) investigate the medical significance of two selected pathology classes by employing visualizations in the form of heatmap score-generated segmentation. In both qualitative and quantitative analyses, we compare the performance of these methods against human expert annotation. To conduct our investigation, we employ the Chexpert dataset, which serves as a suitable resource for the pathology prediction task at hand. Additionally, we utilize the CheXlocalize dataset, a complementary dataset to Chexpert, where human experts have provided annotations for the ground truth values, to facilitate our interpretability investigation.

The field of deep learning offers a wide range of interpretability methods \cite{e23010018, samek2019explainable} tailored to diverse prediction tasks and neural network models. Our research aimed to incorporate both qualitative investigation and quantitative evaluation, necessitating a careful selection of interpretability methods to manage the scope of visualization. One of the chosen methods, LIME, was deemed valuable for interpreting individual instance predictions in black box machine learning models, as it obviates the need for direct manipulation of the original model. Instead, a surrogate model with global interpretability is employed, rendering LIME an attractive option provided that its performance meets satisfactory standards. SHAP, on the other hand, can be viewed as an integration of LIME and Shapley values. By utilizing Shapley values, it quantifies the contribution of each feature, thereby discerning their significance with respect to specific predictions. The foundation of SHAP in game theory lends it theoretical credibility and opens avenues for its potential applicability as supporting evidence in ethical or legal contexts. Consequently, its explanation may be well-received if it aligns with both medical significance and mathematical rigor perspectives. In the context of convolutional neural network (CNN) models, saliency methods \cite{simonyan2014deep, tjoa2022quantifying} hold promise, making Grad-CAM and LRP suitable representatives of this category. It is important to acknowledge that our selection of methods is not exhaustive, as there exist numerous other equally strong candidates. Our future work aims to address those additional methods and expand the scope of investigation accordingly.

Expanding upon the foundations laid in our prior research \cite{9866967}, we have maintained methodological consistency by employing the pre-trained model established in our earlier work. The choice of utilizing a DenseNet-121 model was deliberate, as its superiority over alternative models such as ResNet152 or Inception-v4 has been substantiated by a recent study \cite{saporta2022benchmarking} with relevance to our research domain. This decision was of paramount importance in facilitating a comprehensive and equitable comparison among the various interpretability methods under examination. Consequently, our primary emphasis resides in the evaluation of pathology prediction outcomes, which we deemed satisfactory within the context of our investigation. The inclusion of the 'Fracture' class in the classification may raise certain concerns, especially due to its absence of positive instances in the evaluation set. As detailed in the 'CheXpert Dataset' section, while the validation set underwent manual annotation, the training set was labeled using a rule-based labeler. This discrepancy influenced our decision to refrain from creating an enhanced evaluation set from the training data. Within the training dataset, instances of the 'Fracture' class constitute only 4\% of positive cases and 1.1\% of negative cases \cite{garbin2021structured}. It is also pertinent to note that in the prediction competition organized by the dataset authority \cite{irvin2019chexpert}, the 'Fracture' class was omitted. Consequently, based on the distribution of instances, this class should be considered less relevant to the dataset.

The quantitative evaluation of interpretability methods offers a reasonably impartial and equitable assessment of their performance to a certain extent. However, the overall quantitative performance falls short of achieving satisfaction when compared to human expert annotation. Notably, all four methods examined failed to identify any medical significance for the \textit{Lung Lesion} class. This outcome can be partly attributed to the fact that the predicted data contained only a solitary positive instance of the \textit{Lung Lesion} class, as represented in Table \ref{tab:chexpert_results}.

Curiously, across all interpretability methods, the highest performing class did not exhibit a correlation with either the AUROC pathology prediction evaluation score or the total number of prediction instances within each class. This observation implies that a simple comparison based on different deep neural network models may not be effective, as our conventional evaluation scores, which typically serve as indicators of model performance, did not exhibit strong correlation in this context.

On the whole, Grad-CAM demonstrated the most balanced mean Intersection over Union (IoU) scores across all pathology classes. In contrast, SHAP yielded the poorest scores, as it failed to identify medical significance across four classes and demonstrated highly inconsistent performance for the remaining classes.

\begin{table}
\caption{Mean hit rate of the most representative point in each chest X-ray image}
\label{tab:min-hit-rate}
\begin{tabular}{|l|l|l|l|l|}
\hline
\textbf{\begin{tabular}[c]{@{}l@{}}Pathology\\ Classes\end{tabular}} & \textbf{LIME} & \textbf{SHAP} & \textbf{\begin{tabular}[c]{@{}l@{}}Grad-\\ CAM\end{tabular}} & \textbf{LRP} \\ \hline
\begin{tabular}[c]{@{}l@{}}Lung\\ Opacity\end{tabular} & 0.089 & 0.079 & 0.024 & 0.346 \\ \hline
\begin{tabular}[c]{@{}l@{}}Support\\ Devices\end{tabular} & 0.271 & 0.094 & 0.037 & 0.381 \\ \hline
Atelectasis & 0.025 & 0.013 & 0.063 & 0.315 \\ \hline
Cardiomegaly & 0.311 & 0.193 & 0.176 & 0.706 \\ \hline
Consolidation & 0.031 & 0.0 & 0.061 & 0.0 \\ \hline
Edema & 0.156 & 0.179 & 0.155 & 0.178 \\ \hline
\begin{tabular}[c]{@{}l@{}}Enlarged\\ Cardiomed.\end{tabular} & 0.092 & 0.0 & 0.164 & 0.137 \\ \hline
Lung Lesion & 0.0 & 0.0 & 0.0 & 0.0 \\ \hline
\begin{tabular}[c]{@{}l@{}}Pleural\\ Effusion\end{tabular} & 0.015 & 0.0 & 0.0 & 0.373 \\ \hline
Pneumothorax & 0.245 & 0.0 & 0.0 & 0.124 \\ \hline
\end{tabular}
\end{table}

It is pertinent to acknowledge that the Intersection over Union (IoU) metric exhibits certain limitations \cite{reinke2023understanding}. Primarily, its applicability diminishes in the context of smaller findings. A discrepancy of merely one pixel between the ground truth and predicted segmentation can precipitate a disproportionate decline in the IoU score, especially when compared to larger findings. Furthermore, IoU quantifies the precise pixel overlap, which, although significant, may not fully encompass the critical aspects of medical imaging. In this domain, identifying the relevant region is often as crucial as pinpointing the exact pixels within that region. In this regard, the Hit Rate metric emerges as a complementary measure to IoU \cite{saporta2022benchmarking}. It evaluates whether the most representative point in the heatmap segmentation resides within the ground truth segmentation mask.

Table \ref{tab:min-hit-rate} presents the mean hit rate scores across all four interpretability methods. The data reveal that, with the exception of LRP, the scores from the other three methods align with the mean IoU scores, albeit with minor discrepancies. Notably, LRP demonstrates significantly superior outcomes. This enhancement can be attributed to the qualitative analysis, which indicates that LRP segmentations encompass a substantial area within the images. Consequently, the probability of capturing the most representative point is markedly higher in this instance.

The Explainability Consistency Score (ECS) \cite{van2020systematic} can serve as an alternative method for quantitative evaluation in this context. The computation of the ECS is a two-step process. Initially, both the expert segmentation and the computer-generated heatmap undergo a discretization process (Step 1). Subsequently, Step 2 involves the computation of the agreement between these two discretized representations. This will be an avenue for our future exploration.

Image classification tasks, such as the well-known Imagenet classification \cite{krizhevsky2012imagenet}, predominantly rely on the identification of object segmentation boundaries within an image to distinguish between different object classes. Consequently, achieving improved classification results in such tasks is relatively easier compared to ascertaining the medical significance of chest radiography images. The medical significance of pathology classifications encompasses more nuanced considerations, and the segmentation boundaries in these images tend to be more compact than those encountered in standard image classification tasks.

In light of the observed disparities, we have adopted a different approach in contrast to our previous work \cite{9866967}. Rather than focusing solely on the analysis of best, worst, and ambiguous predictions along with their correct and incorrect variants, we have designed a scenario that offers a fairer evaluation. Specifically, we have selected two pathology classes, namely \textit{Lung Opacity} and \textit{Support Devices}, for in-depth exploration due to their two distinctive characteristics.

Firstly, these classes represent the top-predicted categories based on prediction scores. The \textit{Lung Opacity} prediction yielded AUROC and AUPRC scores of 0.87 and 0.88, respectively. Similarly, the \textit{Support Devices} prediction achieved AUROC and AUPRC scores of 0.89 and 0.87, respectively. Regrettably, as observed during the quantitative evaluation, the highest prediction scores did not align well with improved interpretability in terms of medical significance.

Secondly, these selected classes exhibit similarities to the image boundary segmentation paradigm. For instance, the presence of \textit{Lung Opacity} should ideally be identified as a distinct segment within the lung area, while \textit{Support Devices} should be recognized as an anomalous shape within the chest radiography image, as discussed in the section 'Medical Significance of the Selected Classes'. Building upon this underlying intuition, we can view these classes as tasks involving the detection of segmentation boundaries, akin to standard image classification. Notably, the \textit{Support Devices} class possesses a relatively well-defined shape, making it intriguing to explore whether confusion arises between these two distinct shapes during model identification. It is essential to emphasize that while there is no predefined or general shape for lung opacity, the shape of \textit{Support Devices} is comparatively more distinguishable.

The LIME heatmap score segmentation visualization emerges as a comprehensive tool for assessing the medical significance of pathology classes. Its qualitative investigation yields the best overall performance. In the case of single-label instances, it demonstrates relatively satisfactory outcomes. However, when examining the multi-label scenario, confusion arises between the selected classes. This observation highlights a significant limitation of LIME, particularly in situations where the definition of the 'right' surrounding-local area is flexible. In such cases, its performance tends to be suboptimal. Previous research has illustrated that slight variations in the proximity of two such 'right' surrounding local areas, as observed in a simulated setting, can result in significantly divergent heatmap score visualizations \cite{alvarez2018towards}. Furthermore, the effectiveness of LIME heavily depends on the chosen sampling technique for the local model, and variations in the sampling approach can yield vastly dissimilar outcomes. Moreover, it is disconcerting to note that the manipulation of LIME's results can be accomplished relatively swiftly \cite{slack2020fooling}. Considering these limitations, LIME may not be the most suitable option when confronted with sensitive subjects, such as the detection of medical significance.

The SHAP heatmap score segmentation method exhibited limited success in identifying instances of medical significance, as it failed to capture the majority of relevant cases. In the few instances where it did identify significant regions, it became apparent that the method also included numerous unrelated regions. This outcome may be attributed to a notable drawback associated with the specific variant of SHAP we employed, namely KernelSHAP, which tends to overlook feature dependence—a critical factor in certain cases. Consequently, we observed instances where SHAP erroneously indicated that the significant region for \textit{Lung Opacity} existed outside the actual lung area. Furthermore, there is a legitimate concern that biases may persist undetected within the SHAP framework \cite{slack2020fooling}.

Grad-CAM is specifically designed for convolutional neural network (CNN) models. Its effectiveness is most apparent when there is a well-defined boundary distinction in an image classification task, as it selectively highlights pixels that have a positive impact on the classification outcome. However, when examining the resulting heatmap score segmentation visualizations, we observe that the range of highlighted pixels is relatively extensive. As a consequence, the important area becomes broader, making it challenging to precisely identify the correct region, especially in cases requiring nuanced classification. It is important to note that pixel attribution, which is the basis of Grad-CAM, has been associated with potential risks \cite{ghorbani2019interpretation} and a lack of reliability \cite{kindermans2019reliability}. Therefore, caution should be exercised when employing Grad-CAM for quantitative explanations rather than relying solely on its qualitative interpretability.

The LRP method yielded the least informative visualizations for investigating medical significance, resulting in overall poor performance. LRP is characterized by its production of highly intricate, noisy, and non-class discriminative heatmap score visualizations \cite{jung2021slrp}. As a consequence, interpreting the heatmap score visualizations becomes increasingly challenging, particularly when attempting to discern intricate class explanations that are specific to particular locations and possess nuanced medical significance.

To enrich the understanding of the qualitative study, the 'Appendix' section presents heatmap score segmentation visualizations corresponding to the lowest Intersection over Union (IoU) scores for each selected class with positive predictions. Examination of these visualizations reveals a notable influence of ground truth segmentation size, with smaller sizes frequently overlooked by the heatmap area.

Further analysis, as exemplified in Figure \ref{lime_worst_Support_Devices} where the prediction probability score is 0.86, and juxtaposed with prior observations indicating a lack of correlation between the model's Area Under the Receiver Operating Characteristic (AUROC) and the evaluative metrics (IoU or hit rate) for model explanations, suggests a potential incongruity. Specifically, the most informative explanations may not align with instances where the model exhibits optimal performance. This realization is of paramount importance and cautions against precipitate conclusions regarding the clinical applicability of the interpretability methods under investigation. This area warrants deeper exploration in future research endeavors.

In the realm of local interpretability methods for ML models, a limitation warrants consideration concerning the assessment of qualitative performance. These methods are designed to provide local explanations, elucidating the model's reasoning for individual samples. While such granularity of explanation can be beneficial, it simultaneously poses a challenge in synthesizing a global perspective of the model's behavior. Local explanations, in their inherent nature, may not always be illuminative in discerning whether the ML model has effectively captured clinically relevant patterns across the entire dataset. Consequently, basing evaluations on a limited selection of examples could potentially lead to a skewed representation of the model's interaction with the full dataset.

This study opts not to assert general conclusions about the overall performance, including the strengths and weaknesses, of these interpretability methods. Instead, it aims to illuminate the disparities between quantitative metrics and qualitative insights, drawing attention to crucial gaps based on the limited scope of the qualitative analysis conducted.

A further limitation of this qualitative evaluation lies in the absence of domain-specific expertise in the appraisal process. The interpretations and explanations produced by the model have not undergone evaluation by domain experts, such as clinicians or radiologists. This gap could raise concerns about the applicability and perceived utility of these explanations in a clinical setting. While comparisons are made between model-generated heatmaps and ground truth segmentations of targeted findings, it is essential to recognize that medical images often contain numerous clinically relevant details that extend beyond the immediate findings under consideration. These additional details could be pivotal for medical practitioners in the diagnostic process.

Moreover, the alignment of model explanations with pre-existing human knowledge might not be a definitive criterion for medical significance or clinical relevance. As highlighted in \cite{jin2023guidelines}, explanations have the potential to uncover new insights. Overemphasis on creating explanations that align with user expectations may inadvertently lead to deceptive explanations that provide seemingly plausible rationales for incorrect model predictions. To mitigate this concern, our future studies will actively involve domain experts, ensuring that the evaluation of model explanations is grounded in relevant clinical expertise and context.

The exploration of medical significance in deep neural network-based predictions is a multifaceted undertaking that demands a thorough examination from both quantitative and qualitative perspectives, as demonstrated in this study. Relying solely on the quantitative aspects of LRP heatmap score segmentation, for instance, would have provided an incomplete understanding of the medical significance, as the qualitative exploration presented a distinct viewpoint. To achieve more robust and informative outcomes for medical diagnosis prediction tasks employing deep neural networks, it is imperative to consider the interplay of three key aspects: (1) the data, (2) the prediction model, and (3) the interpretability method.

Firstly, it is crucial to recognize that an interpretability method is inherently dependent on the underlying model. Hence, efforts should be directed towards improving the model's alignment with real-world aspects. Enhancing the model's ability to emulate reality increases the transferability and reliability of the model's interpretations. In this context, the integration of multimodal data, encompassing not only chest radiography but also patient history, clinical notes, and laboratory tests, can be instrumental. By leveraging multimodal information, the explanations for predictions can be mutually reinforcing, potentially enhancing the interpretability of the results.

Secondly, infusing domain knowledge into the model architecture is of significant value. Incorporating multimodal data is one approach to achieving this goal. Exploring different model architectures is another avenue worth exploring. While CNNs are well-suited for image data as they capture spatial dependencies effectively, our findings indicate that their performance was unsatisfactory in this particular context. Vision transformers \cite{dosovitskiy2020vit, touvron2021training}, on the other hand, have demonstrated superior performance and should be investigated \cite{chefer2021transformer} further. To leverage the benefits of multimodality, employing techniques such as ClinicalBERT \cite{alsentzer-etal-2019-publicly}, along with variations of ImageBERT \cite{qi2020imagebert}, or Pixel-BERT \cite{huang2020pixel} fine-tuned on chest radiography data and tailored to the specific diagnosis task(s), could yield promising outcomes.

Lastly, the implementation and refinement of interpretability methods should be guided by causal considerations (if feasible), feature dependence, and a deeper understanding of the landscape of chest radiography or medical data in general. The relevant regions within these chest radiography images are often in close proximity with frequent overlaps, and the nuances of their interpretations should be adequately reflected in the methods employed.

The present study differentiates itself from the body of related works, as outlined in the section 'Related Works', primarily in its approach and scope. A common trend observed in most related studies is their inclination towards illustrating a few select examples as best-case scenarios to validate their interpretability methods. This is often done without engaging in a systematic comparison of the same instances across different methods. An exception to this trend is found in \cite{9576766}, where a comparative analysis between LIME and SHAP is presented. Furthermore, only a handful of studies, notably \cite{van2020systematic} and \cite{saporta2022benchmarking}, have engaged in a quantitative investigation, specifically focusing on saliency methods.

It is imperative to note that the primary objective of our study diverges from benchmarking performance. Instead, our endeavor is to provide a comprehensive overview concerning the application of these interpretability methods. In doing so, we aim to underscore the inherent limitations associated with such analysis approaches. This exploration is grounded in the theoretical foundations of the interpretability methods under scrutiny. Additionally, we endeavor to offer insights that are derived not only from the theoretical underpinnings but also from the performance attributes of these methods.

The challenges encountered in this study underscore the pressing need for continued research and advancement within this domain. In response to these challenges, our future research will be oriented towards adopting a multi-modality-based approach. This approach entails the integration of multiple information sources, utilization of diverse modalities, and exploration of various model architectures. Our objective is to significantly improve the understanding and interpretation of medical significance inherent in deep neural network predictions.

Moreover, we plan to explore different relevant quantitative assessment methods to facilitate a more comprehensive comparison of these interpretability methods. In addition, we intend to develop a robust framework that incorporates impactful qualitative analysis, augmented by the active guidance of medical experts.

It is our sincere hope that this investigative pathway will prove to be substantially promising in enhancing the reliability and practicality of these predictions within the sphere of medical diagnostics and decision-making.

\section{Conclusion}

This study undertakes a comprehensive examination of four prominent interpretability methods: LIME (Local Interpretable Model-agnostic Explanations), SHAP (Shapley Additive exPlanations), Grad-CAM (Gradient-weighted Class Activation Mapping), and LRP (Layer-wise Relevance Propagation). The principal aim is to critically assess the medical significance embedded within a deep learning-based pathology prediction model, particularly applied to multi-label chest radiography data. A meticulous analysis, comprising both quantitative and qualitative assessments in comparison with human expert annotation, is conducted to provide a holistic and realistic evaluation of the current capabilities in discerning medical significance. Quantitative evaluation indicates that Grad-CAM exhibits relatively superior performance. Conversely, qualitative analysis reveals that LIME heatmap score segmentation visualization demonstrates enhanced medical significance. This research endeavors to shed light on both the findings and limitations encountered in the holistic approach employed for evaluating these interpretability methods. The intent is to contribute to the advancement of their practical application in the realm of medical diagnosis, offering insights and charting future directions. The findings from this study are intended to serve as a valuable resource for researchers, practitioners, and stakeholders striving to augment the interpretability and reliability of deep learning models in the vital domain of medical diagnostics. Moreover, we acknowledge the nascent state of this field and advocate for continued exploration and refinement of interpretable methods and evaluation techniques. Such endeavors are crucial to propel ongoing progress in the sector of medical diagnosis.

\section{Conflict of interest}
Authors state no conflict of interest.

\section{Resource Availability}

The resource can be obtained using the following links,

\begin{enumerate}

\item \href{https://github.com/anondo1969/SHAMSUL}{\underline{The Code Implementation}}

\item \href{https://github.com/anondo1969/SHAMSUL/tree/main/codes/Model_Weights}{\underline{The Trained Model}}

\item \href{https://stanfordmlgroup.github.io/competitions/chexpert/}{\underline{The CheXpert Dataset}}

\item \href{https://github.com/rajpurkarlab/cheXlocalize/blob/master/download_instructions.md}{\underline{The CheXlocalize Dataset}}

\end{enumerate}

\section{Appendix}

\begin{figure}[ht]
    \centering
    \begin{subfigure}{.18\textwidth}
        \centering
        \includegraphics[width=1\linewidth]{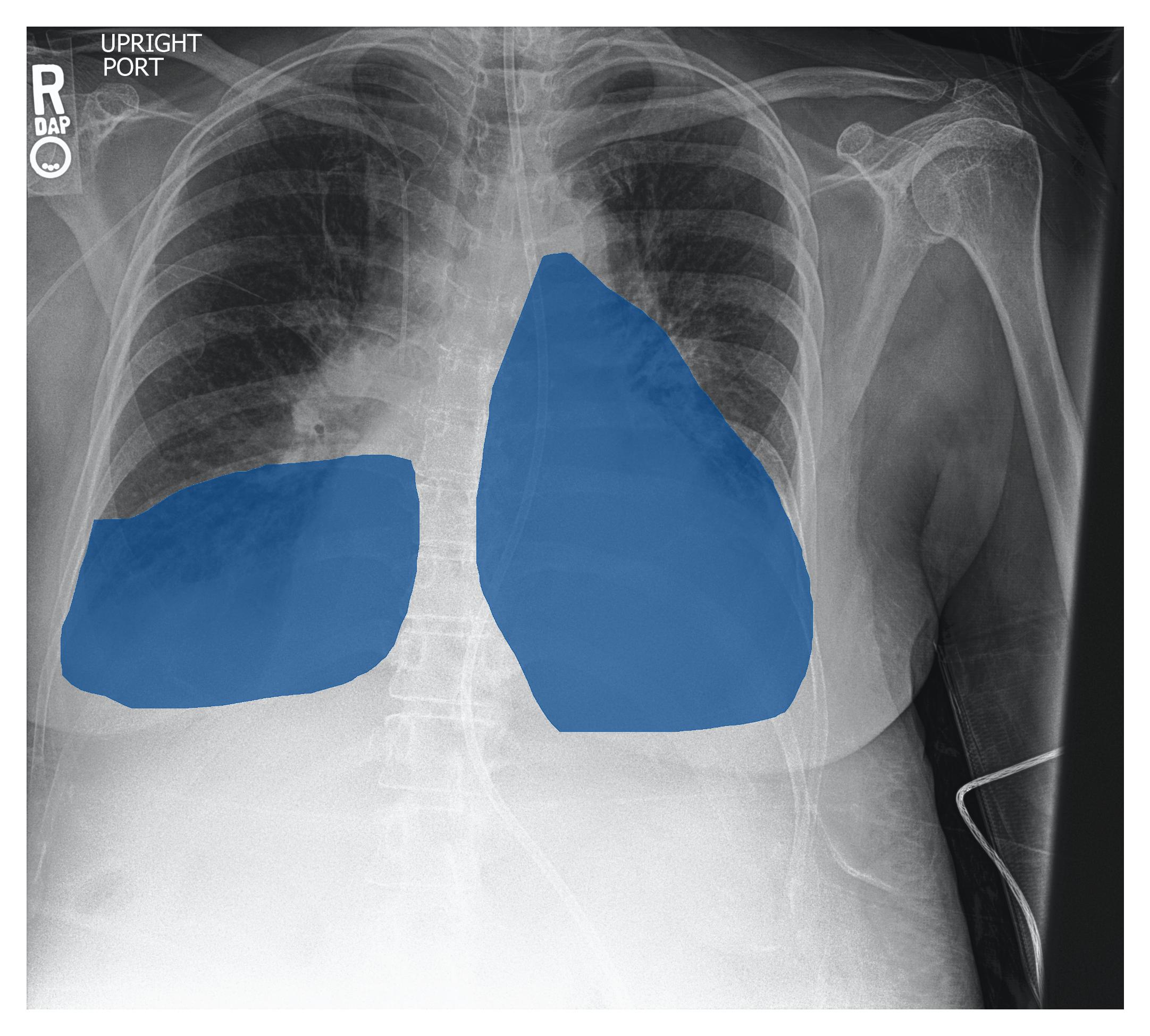}
        \caption{Ground Truth}
        \label{worst_gt_lime_Lung_Opacity}
    \end{subfigure} 
    \begin{subfigure}{.18\textwidth}
        \centering
        \includegraphics[width=1\linewidth]{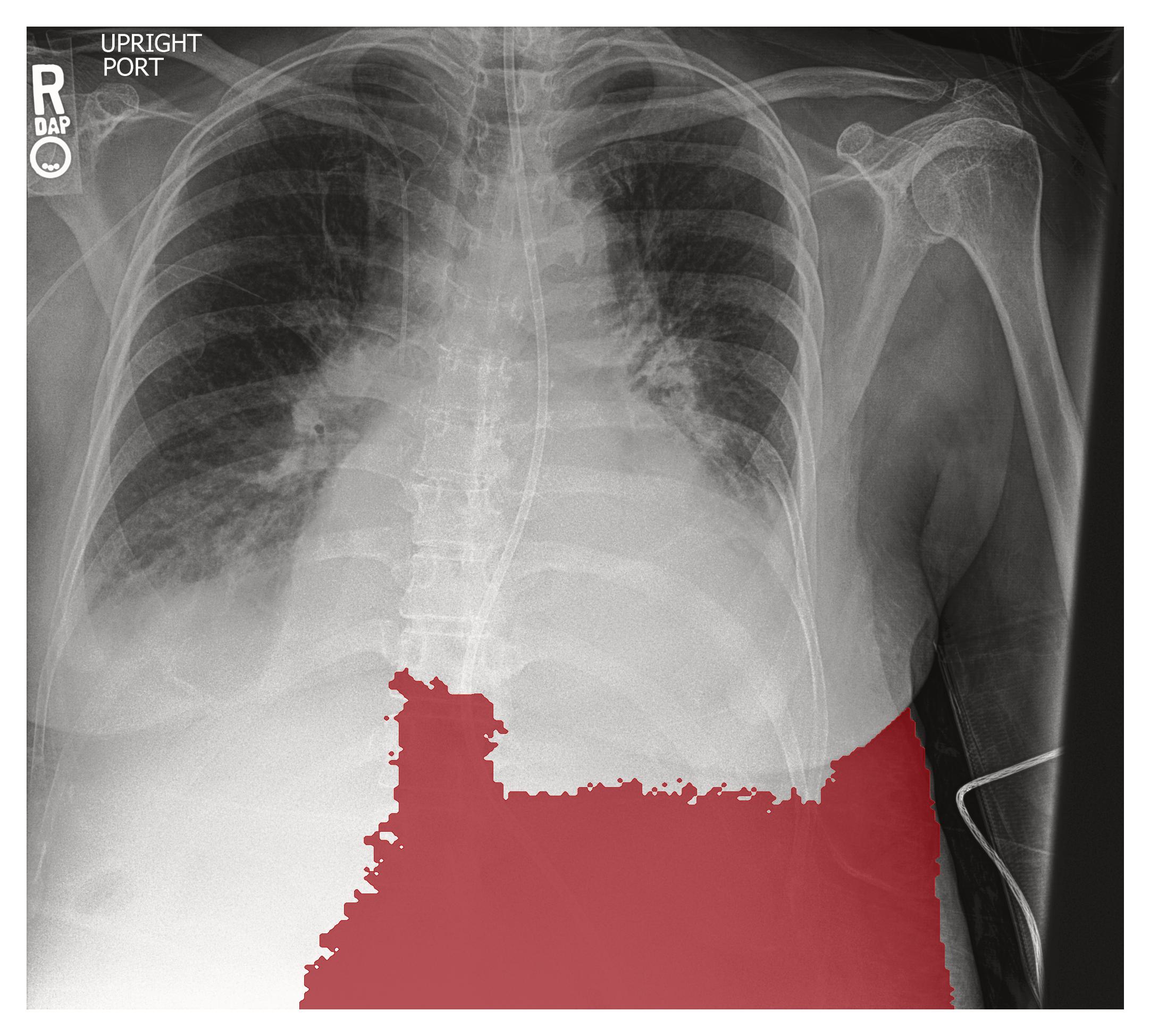}
        \caption{LIME Segment.}
        \label{worst_pred_lime_Lung_Opacity}
    \end{subfigure}

    \caption{\textbf{LIME} heatmap score segmentation of the \textit{\textbf{Lung Opacity}} (IoU score: 0.0, prediction probability score: 0.54) class with human expert segmentation annotation for comparison.}
    
    \label{lime_worst_Lung_Opacity}
\end{figure}

\begin{figure}[ht]
    \centering
    \begin{subfigure}{.18\textwidth}
        \centering
        \includegraphics[width=1\linewidth]{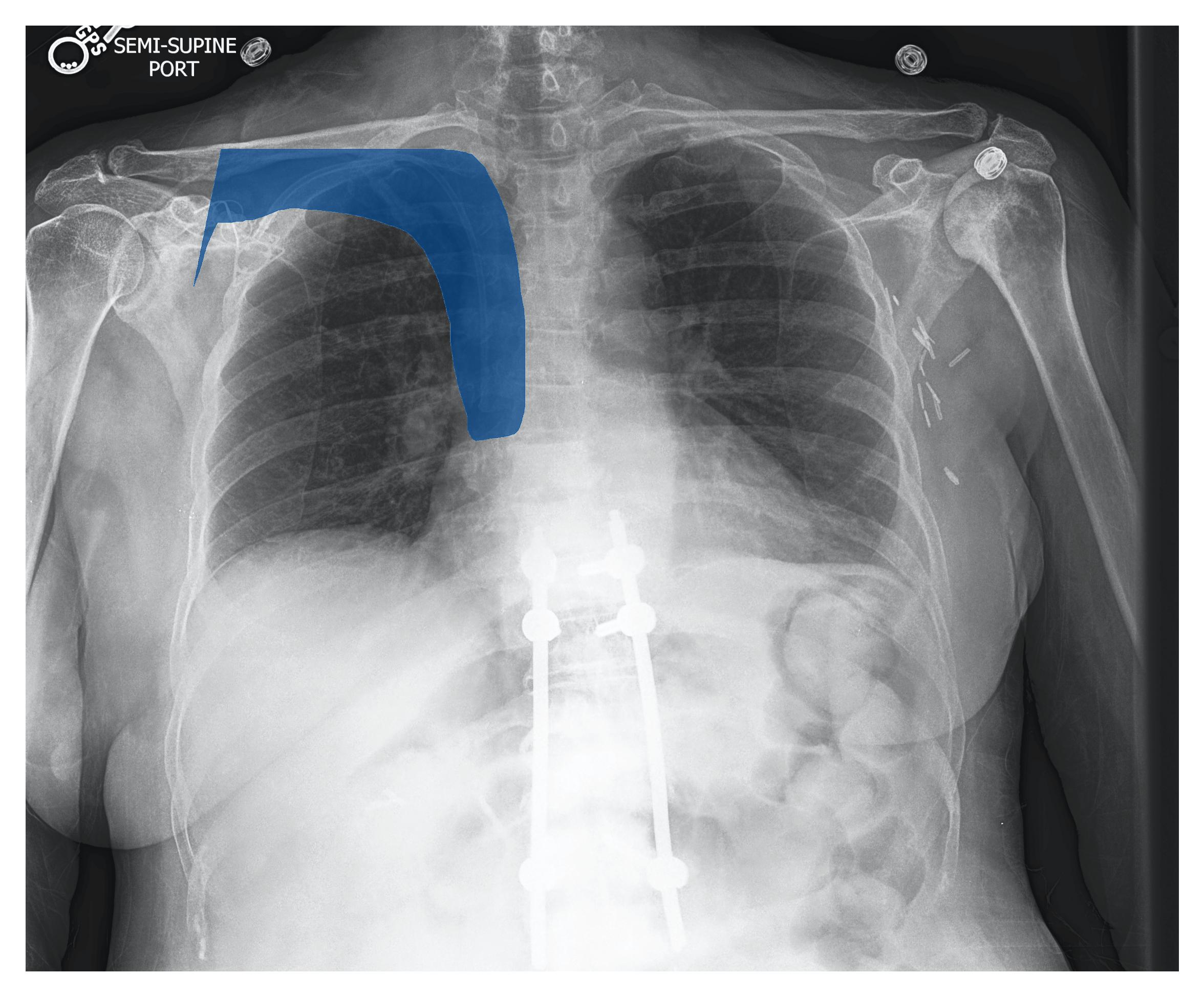}
        \caption{Ground Truth}
        \label{worst_gt_lime_Support_Devices}
    \end{subfigure} 
    \begin{subfigure}{.18\textwidth}
        \centering
        \includegraphics[width=1\linewidth]{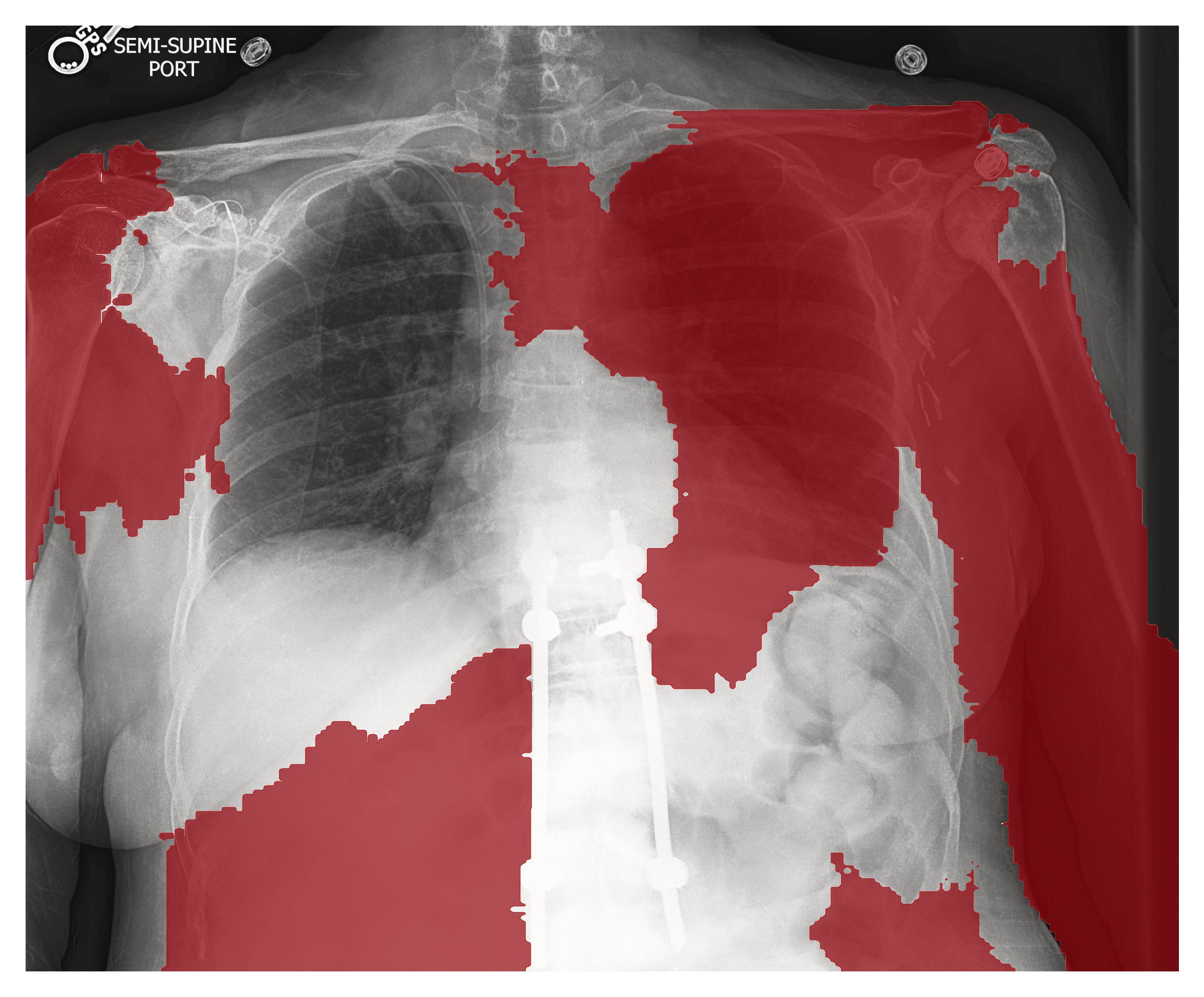}
        \caption{LIME Segment.}
        \label{worst_pred_lime_Support_Devices}
    \end{subfigure}

    \caption{\textbf{LIME} heatmap score segmentation of the \textit{\textbf{Support Devices}} (IoU score: 0.004, prediction probability score: 0.86) class with human expert segmentation annotation for comparison.}
    
    \label{lime_worst_Support_Devices}
\end{figure}

\begin{figure}[ht]
    \centering
    \begin{subfigure}{.18\textwidth}
        \centering
        \includegraphics[width=1\linewidth]{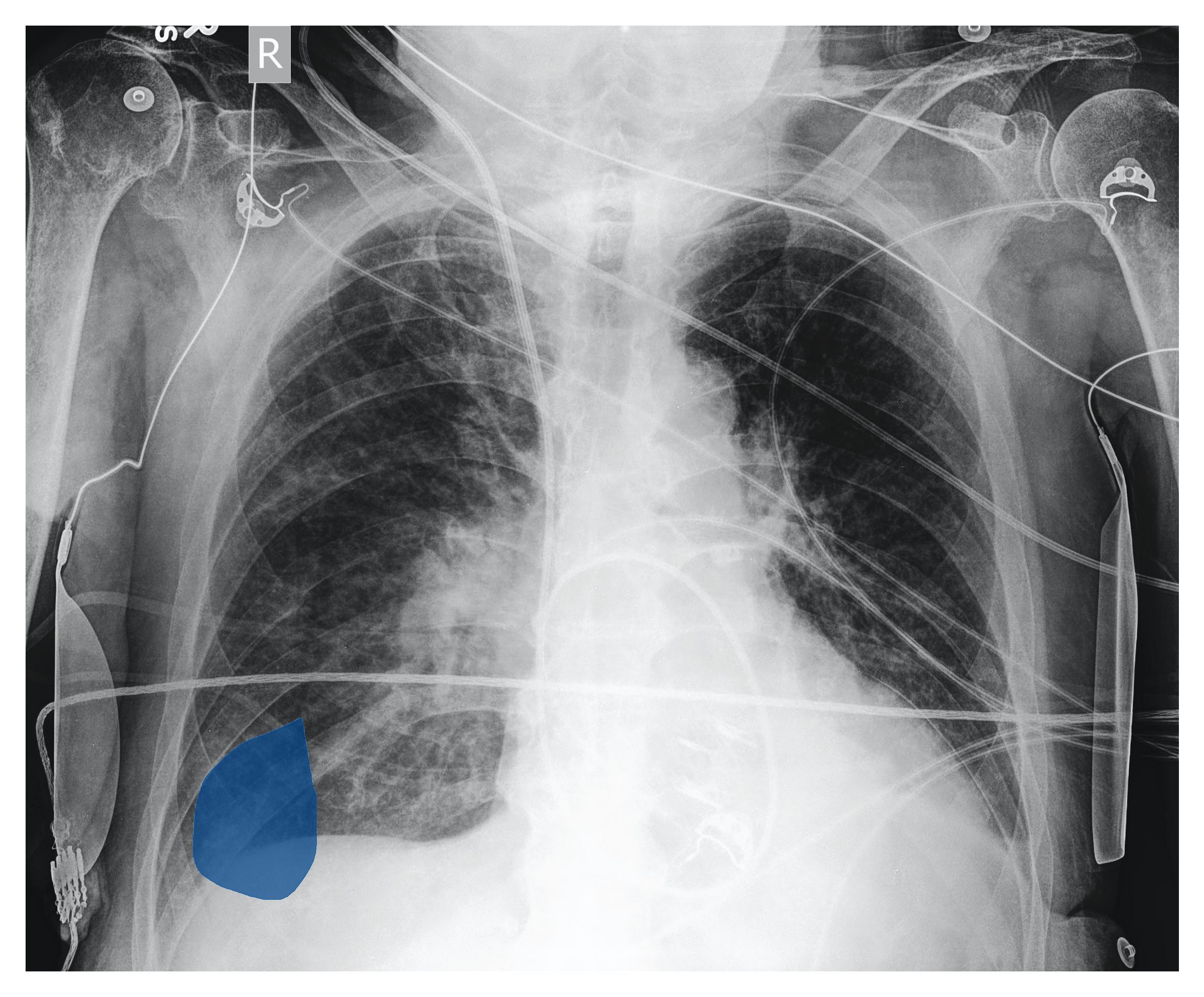}
        \caption{Ground Truth}
        \label{worst_gt_shap_Lung_Opacity}
    \end{subfigure} 
    \begin{subfigure}{.18\textwidth}
        \centering
        \includegraphics[width=1\linewidth]{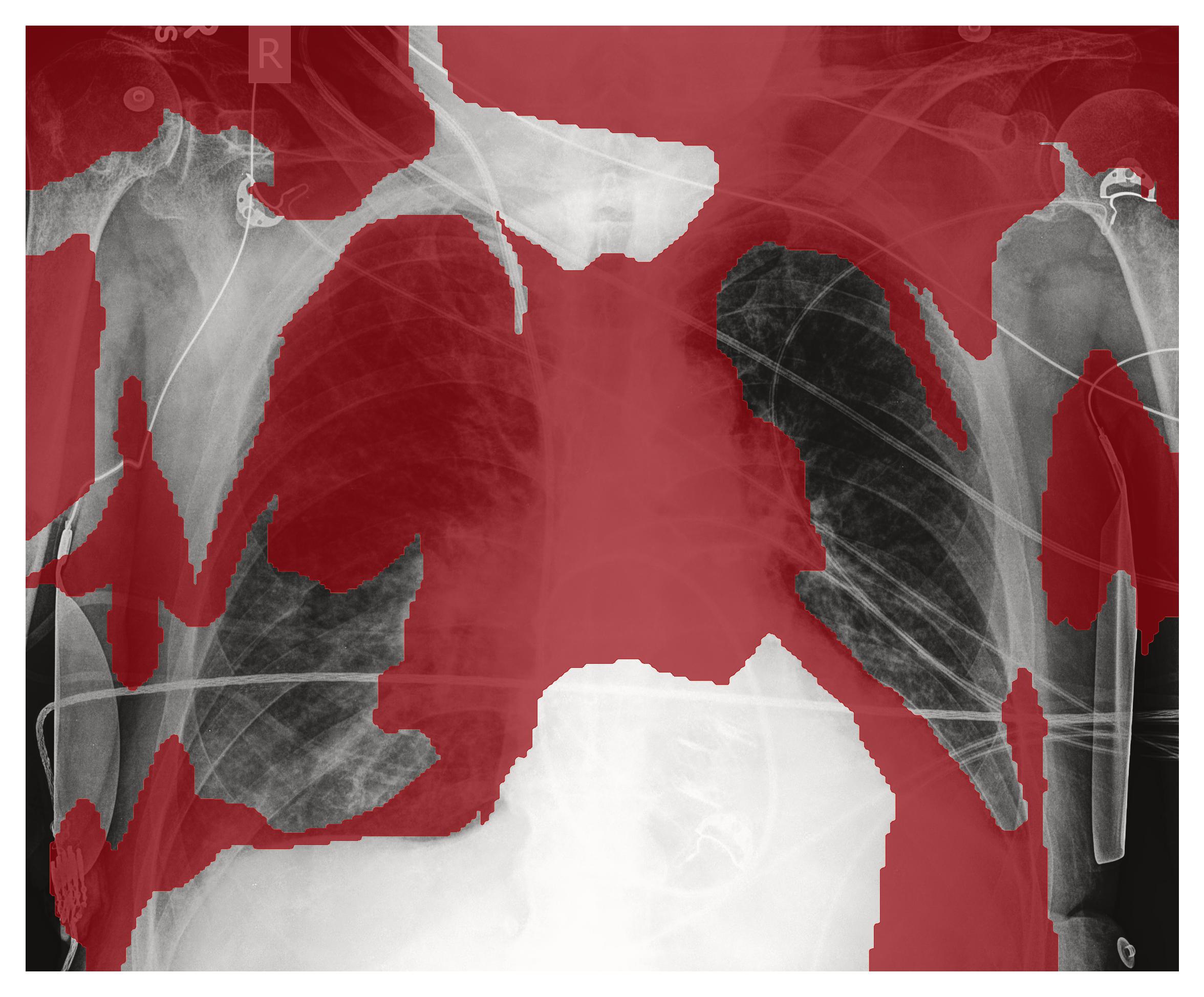}
        \caption{SHAP Segment.}
        \label{worst_pred_shap_Lung_Opacity}
    \end{subfigure}

    \caption{\textbf{SHAP} heatmap score segmentation of the \textit{\textbf{Lung Opacity}} (IoU score: 0.008, prediction probability score: 0.60) class with human expert segmentation annotation for comparison.}
    
    \label{shap_worst_Lung_Opacity}
\end{figure}

\begin{figure}[ht]
    \centering
    \begin{subfigure}{.18\textwidth}
        \centering
        \includegraphics[width=1\linewidth]{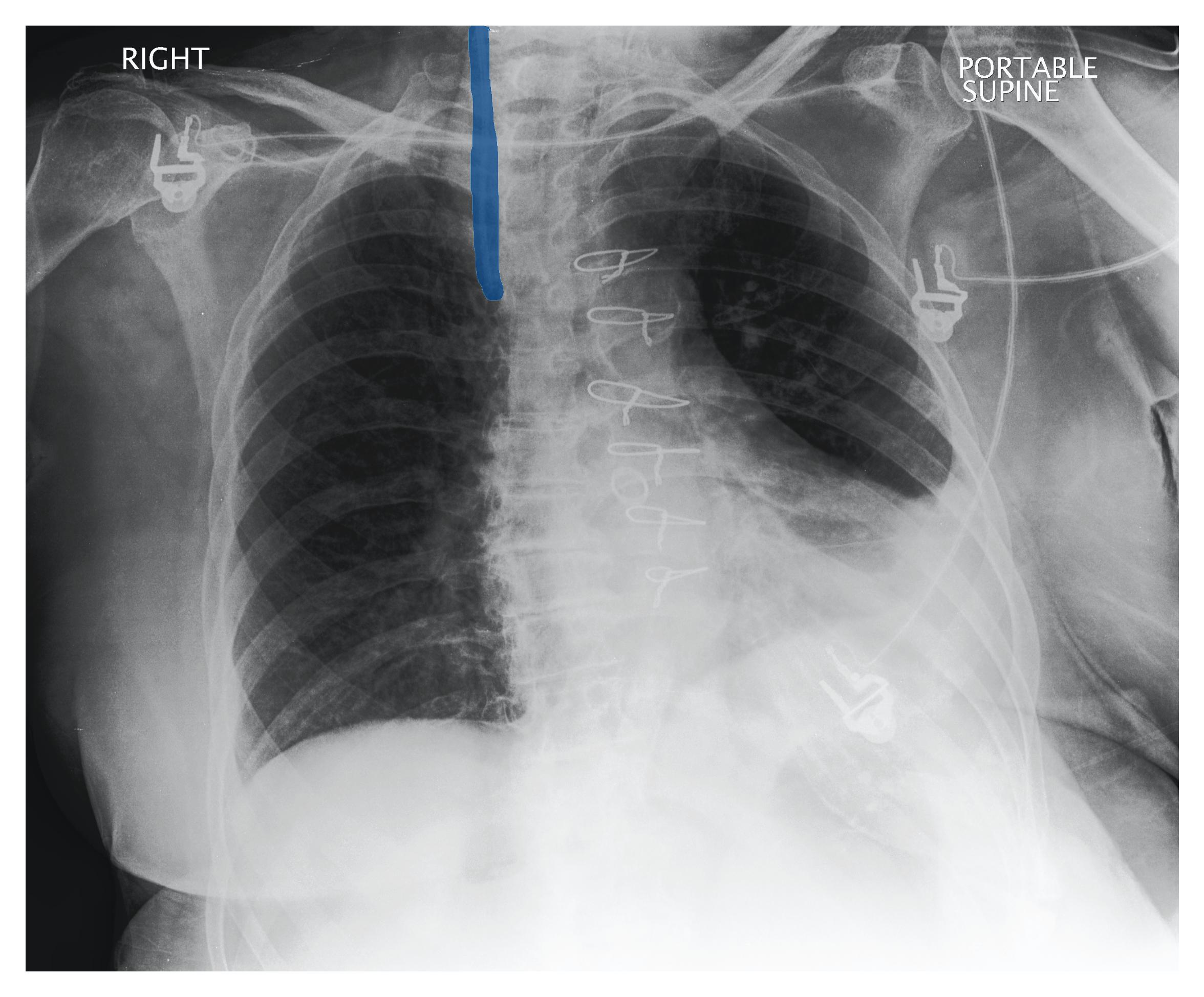}
        \caption{Ground Truth}
        \label{worst_gt_shap_Support_Devices}
    \end{subfigure} 
    \begin{subfigure}{.18\textwidth}
        \centering
        \includegraphics[width=1\linewidth]{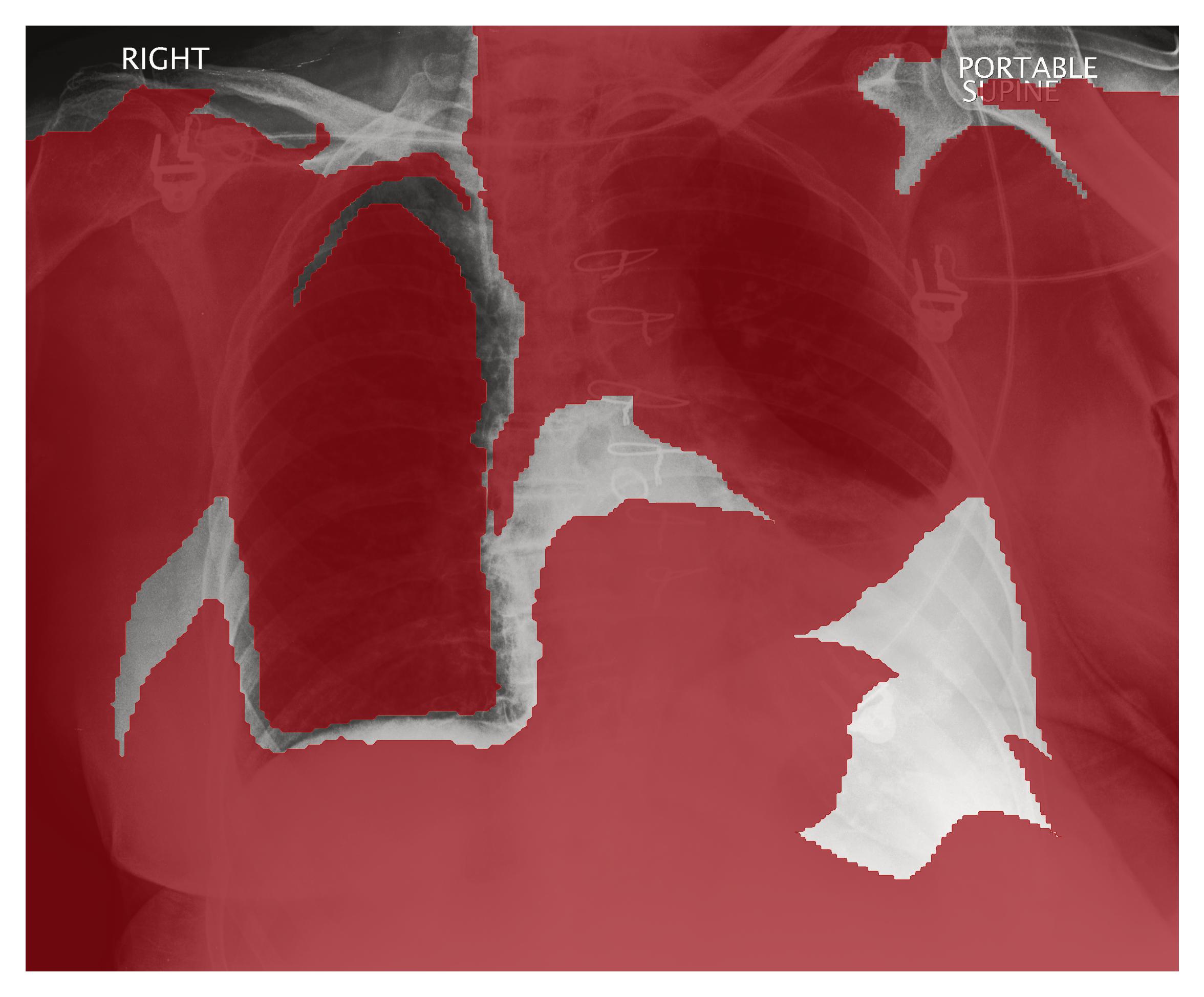}
        \caption{SHAP Segment.}
        \label{worst_pred_shap_Support_Devices}
    \end{subfigure}

    \caption{\textbf{SHAP} heatmap score segmentation of the \textit{\textbf{Support Devices}} (IoU score: 0.003, prediction probability score: 0.55) class with human expert segmentation annotation for comparison.}
    
    \label{shap_worst_Support_Devices}
\end{figure}

\begin{figure}[ht]
    \centering
    \begin{subfigure}{.18\textwidth}
        \centering
        \includegraphics[width=1\linewidth]{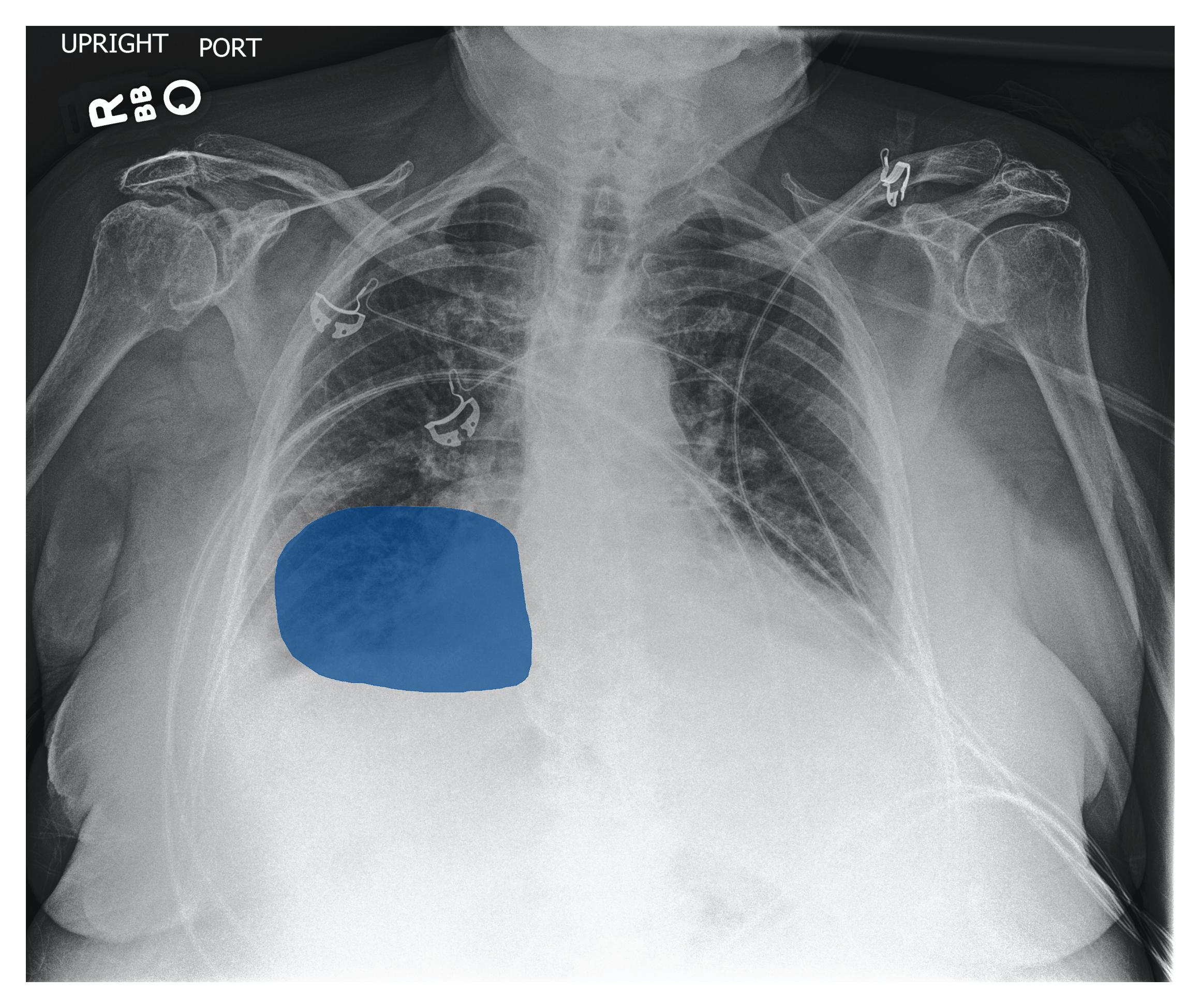}
        \caption{Ground Truth}
        \label{worst_gt_grad_cam_Lung_Opacity}
    \end{subfigure} 
    \begin{subfigure}{.18\textwidth}
        \centering
        \includegraphics[width=1\linewidth]{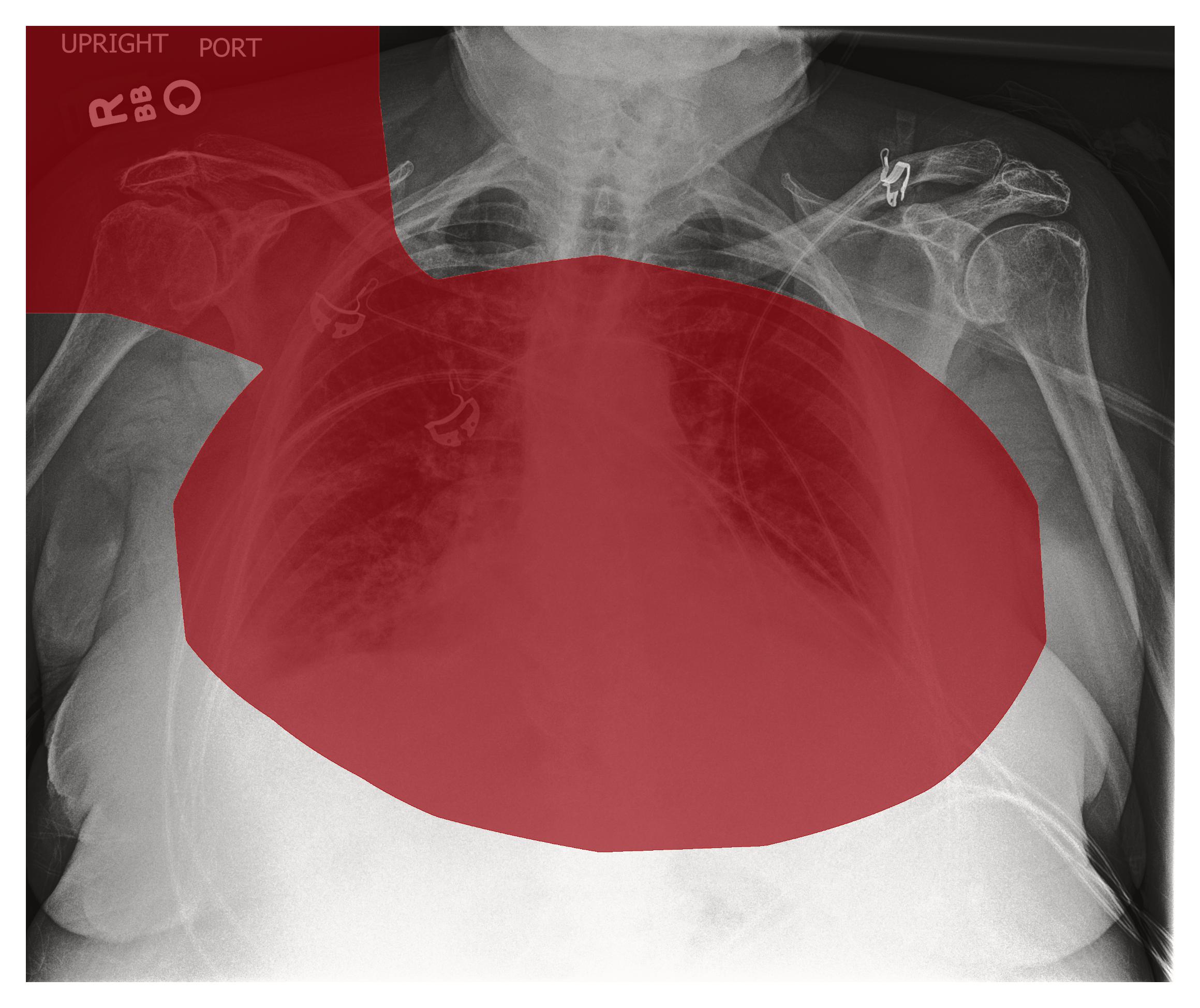}
        \caption{Grad-CAM Seg.}
        \label{worst_pred_grad_cam_Lung_Opacity}
    \end{subfigure}

    \caption{\textbf{Grad-CAM} heatmap score segmentation of the \textit{\textbf{Lung Opacity}} (IoU score: 0.08, prediction probability score: 0.53) class with human expert segmentation annotation for comparison.}
    
    \label{grad_cam_worst_Lung_Opacity}
\end{figure}

\begin{figure}[ht]
    \centering
    \begin{subfigure}{.18\textwidth}
        \centering
        \includegraphics[width=1\linewidth]{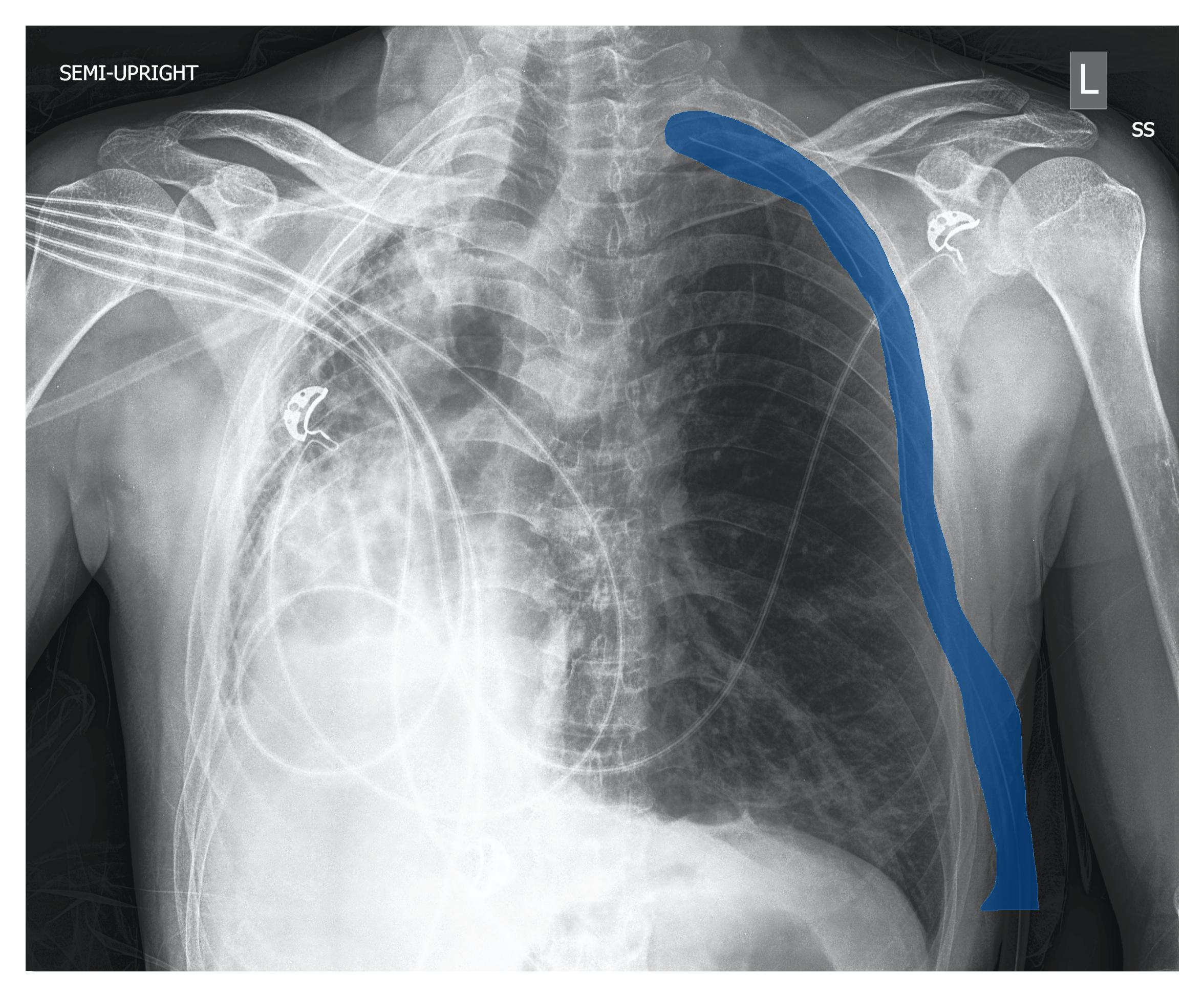}
        \caption{Ground Truth}
        \label{worst_gt_grad_cam_Support_Devices}
    \end{subfigure} 
    \begin{subfigure}{.18\textwidth}
        \centering
        \includegraphics[width=1\linewidth]{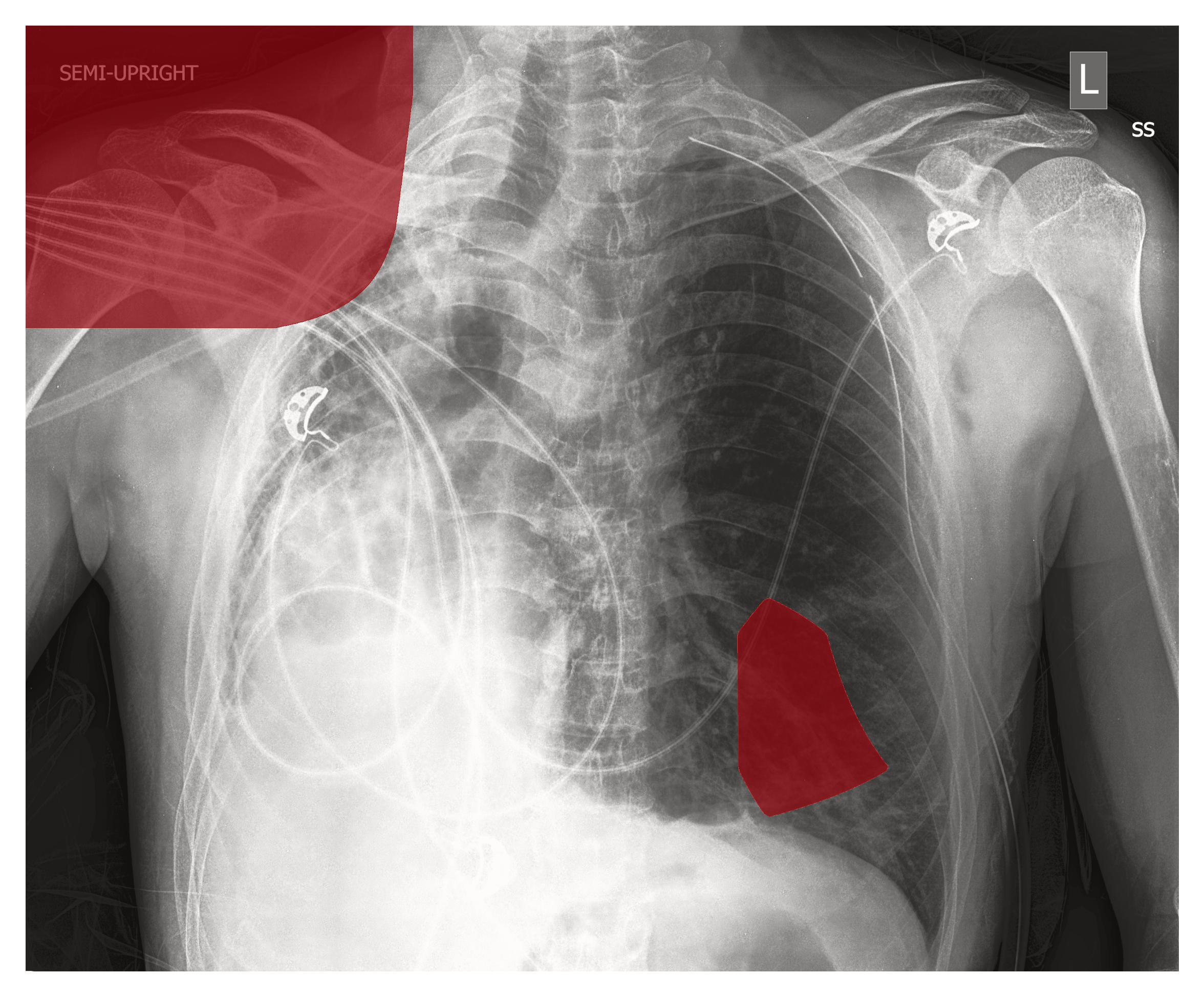}
        \caption{Grad-CAM Seg.}
        \label{worst_pred_grad_cam_Support_Devices}
    \end{subfigure}

    \caption{\textbf{Grad-CAM} heatmap score segmentation of the \textit{\textbf{Support Devices}} (IoU score: 0.01, prediction probability score: 0.67) class with human expert segmentation annotation for comparison.}
    
    \label{grad_cam_worst_Support_Devices}
\end{figure}

\begin{figure}[ht]
    \centering
    \begin{subfigure}{.18\textwidth}
        \centering
        \includegraphics[width=1\linewidth]{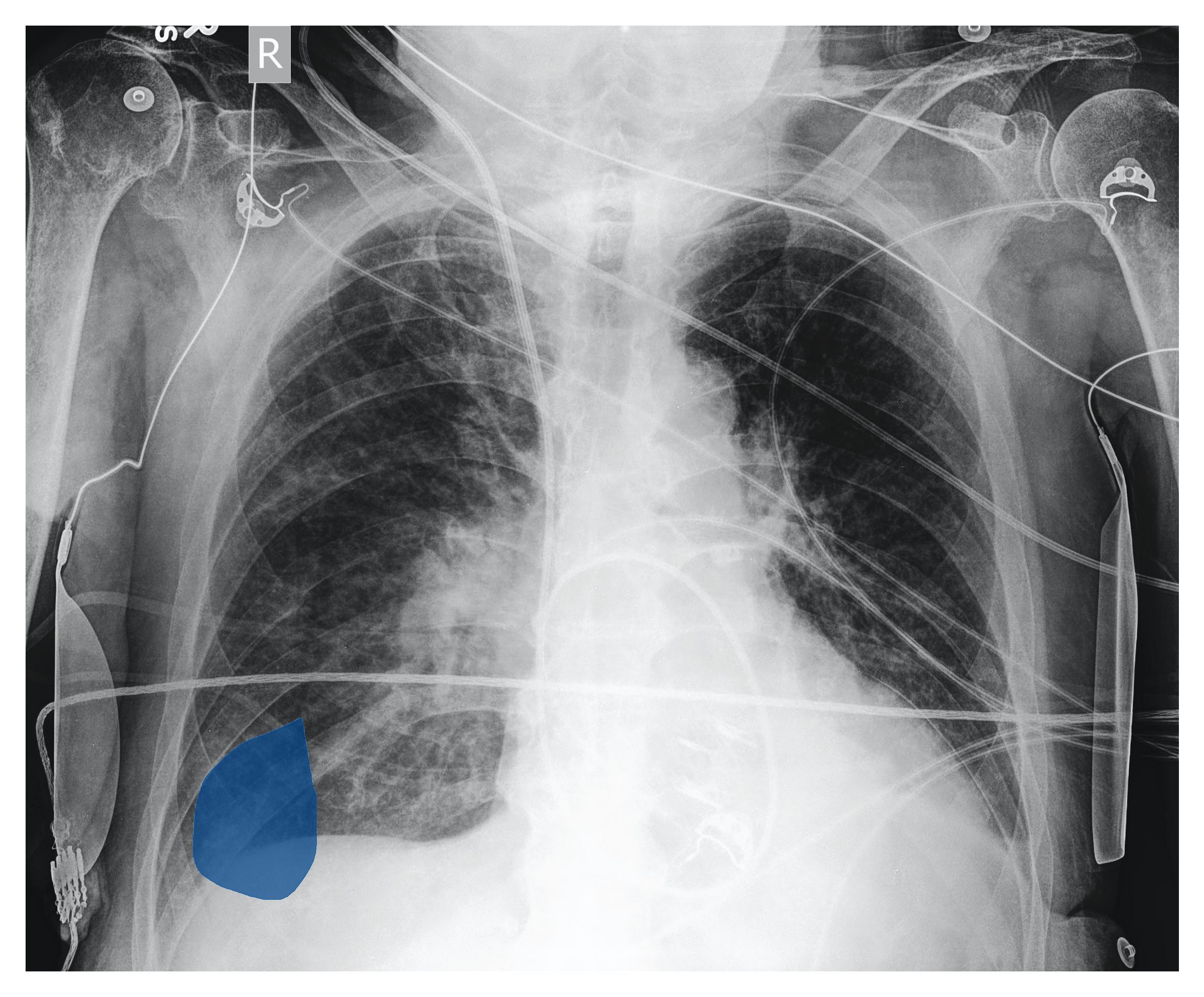}
        \caption{Ground Truth}
        \label{worst_gt_lrp_Lung_Opacity}
    \end{subfigure} 
    \begin{subfigure}{.18\textwidth}
        \centering
        \includegraphics[width=1\linewidth]{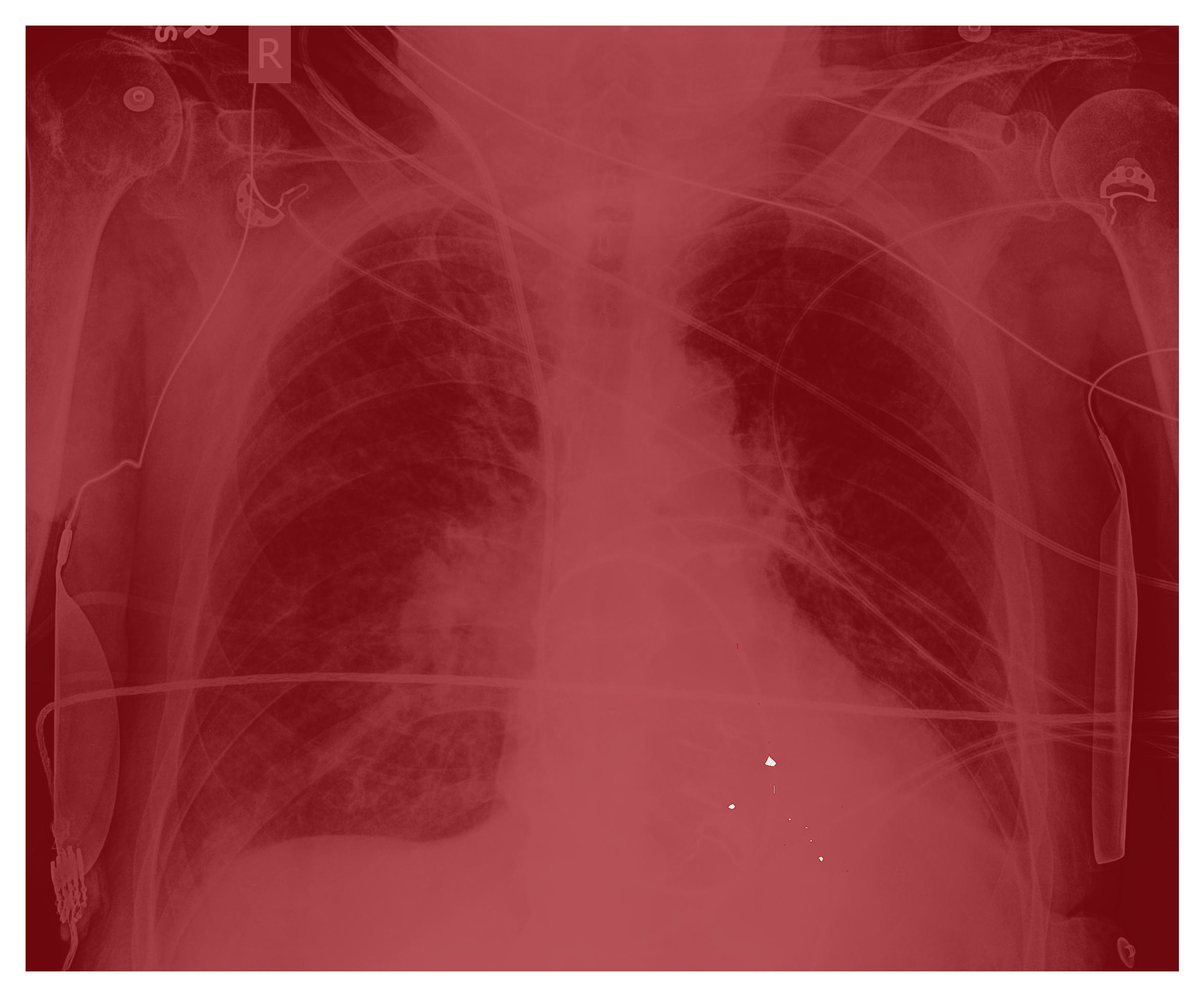}
        \caption{LRP Segment.}
        \label{worst_pred_lrp_Lung_Opacity}
    \end{subfigure}

    \caption{\textbf{LRP} heatmap score segmentation of the \textit{\textbf{Lung Opacity}} (IoU score: 0.02, prediction probability score: 0.60) class with human expert segmentation annotation for comparison.}
    
    \label{lrp_worst_Lung_Opacity}
\end{figure}

\begin{figure}[ht]
    \centering
    \begin{subfigure}{.18\textwidth}
        \centering
        \includegraphics[width=1\linewidth]{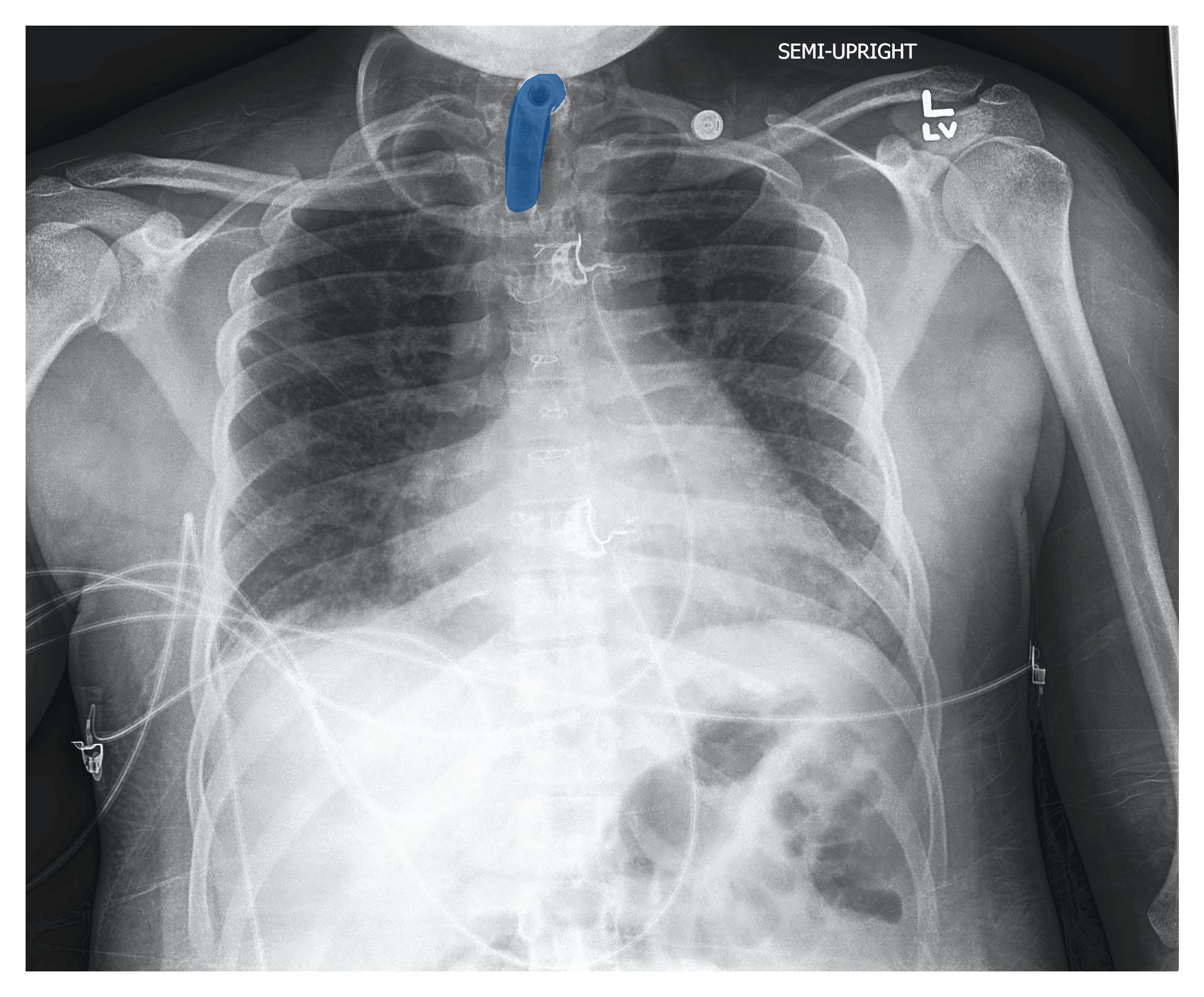}
        \caption{Ground Truth}
        \label{worst_gt_lrp_Support_Devices}
    \end{subfigure} 
    \begin{subfigure}{.18\textwidth}
        \centering
        \includegraphics[width=1\linewidth]{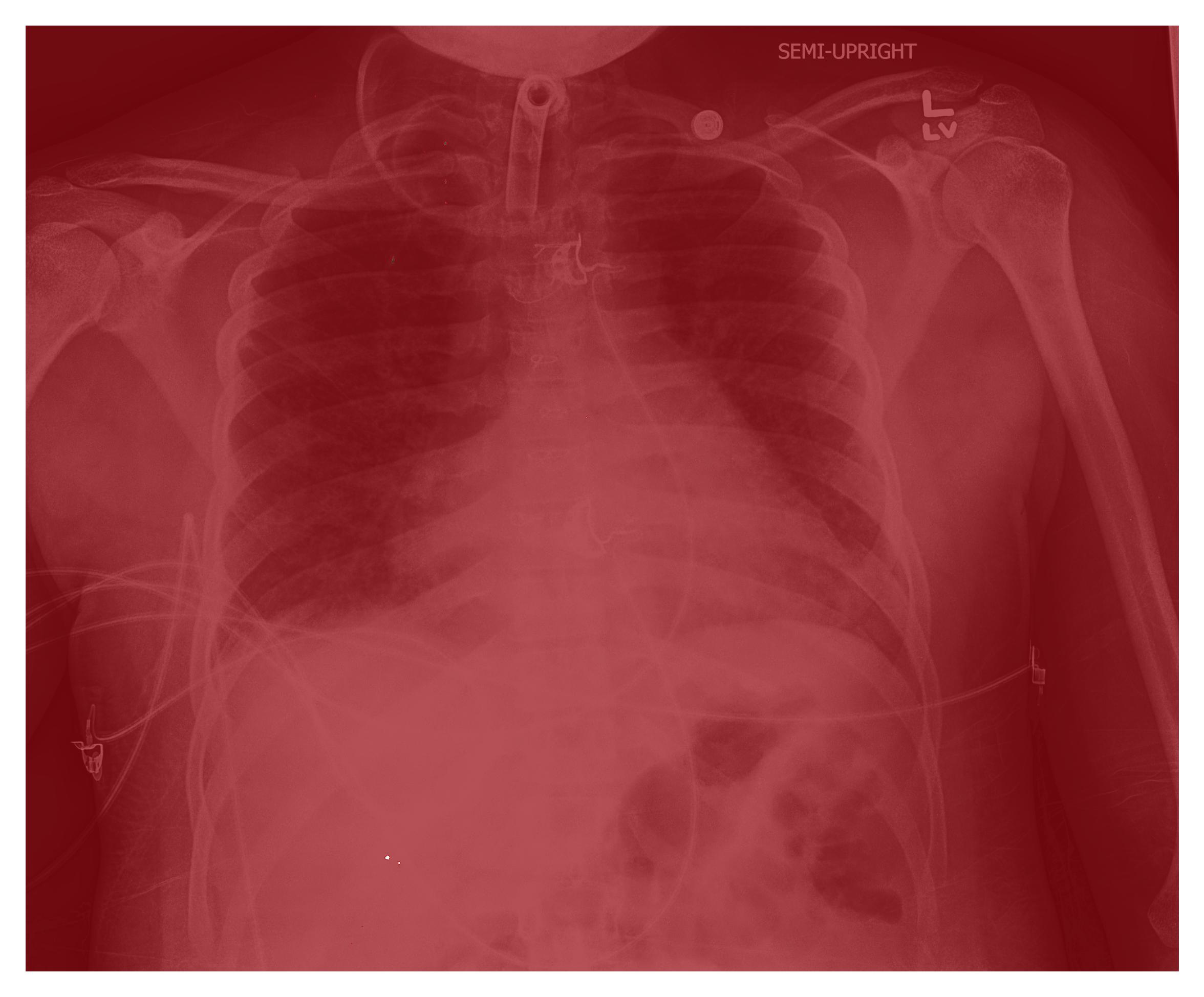}
        \caption{LRP Segment.}
        \label{worst_pred_lrp_Support_Devices}
    \end{subfigure}

    \caption{\textbf{LRP} heatmap score segmentation of the \textit{\textbf{Support Devices}} (IoU score: 0.005, prediction probability score: 0.70) class with human expert segmentation annotation for comparison.}
    
    \label{lrp_worst_Support_Devices}
\end{figure}

The subsequent figures (\ref{lime_worst_Lung_Opacity} to \ref{lrp_worst_Support_Devices}) display heatmap score segmentation visualizations corresponding to the lowest Intersection over Union (IoU) score for each of the selected classes in cases where the predictions were positive.

\cleardoublepage

\bibliographystyle{unsrtnat}
\bibliography{main.bib}

\end{document}